\documentclass[prx, aps, reprint, longbibliography] {revtex4-1}
\usepackage{times, mathrsfs, amsmath, amsfonts, graphics, graphicx, cancel, color, theorem, bbm, mathtools, amssymb}

\usepackage[english]{babel}
\usepackage[margin=.75in]{geometry}
\usepackage[T1]{fontenc}
\usepackage{xfrac,relsize,float}
\usepackage[unicode=true, breaklinks=false, pdfborder={0 0 1}, backref=false, colorlinks=true, linkcolor=blue, citecolor=blue]{hyperref}
\usepackage{physics}

\definecolor{Blue}{rgb}{0,0,1}
\definecolor{Red}{rgb}{1,0,0}
\definecolor{Green}{rgb}{0,1,0}
\definecolor{darkgreen}{rgb}{0,.7,0}
\definecolor{Purp}{rgb}{.2,0,.2}
\definecolor{white}{rgb}{1,1,1}

\newcommand{\figref}[1]{Fig.~\ref{#1}}

\newcommand{\secref}[1]{Sec.~\ref{#1}}
\newcommand{\appref}[1]{App.~\ref{#1}}

\newcommand{\Id}{\mathbbm{1}}

\usepackage[most]{tcolorbox}

\begin{document}
\title{ Tight Bound on Finite-Resolution Quantum Thermometry at Low Temperatures }
\author{Mathias R. J{\o}rgensen${}^{1}$}
    \email{matrj@fysik.dtu.dk}
\author{Patrick P. Potts${}^{2}$}
\author{Matteo G. A. Paris${}^{3}$}
\author{and Jonatan B. Brask${}^{1}$}
\affiliation{${}^{1}$Department of Physics, Technical University of Denmark, 2800 Kongens Lyngby, Denmark}
\affiliation{${}^{2}$Physics Department and NanoLund, Lund University, Box 118, 22100 Lund, Sweden}
\affiliation{${}^{3}$Quantum Technology Lab, Dipartimento di Fisica $"$Aldo Pontremoli$"$, Universit\`a degli Studi di Milano, I-20133 Milano, Italy}
\date{\today}

\begin{abstract}
Precise thermometry is of wide importance in science and technology in general and in quantum systems in particular. Here, we investigate fundamental precision limits for thermometry on cold quantum systems,
taking into account constraints due to finite measurement resolution. We derive a tight bound on the optimal precision scaling with temperature, as the temperature approaches zero. The bound demonstrates that under finite resolution, the variance in any temperature estimate must decrease slower than linearly. This scaling can be saturated by monitoring the non-equilibrium dynamics of a single-qubit probe. We support this finding by numerical simulations of a spin-boson model. In particular, this shows that thermometry with a vanishing absolute error at low temperature is possible with finite resolution, answering an interesting question left open by previous work. Our results are relevant both fundamentally, as they illuminate the ultimate limits to quantum thermometry, and practically, in guiding the development of sensitive thermometric techniques applicable at ultracold temperatures.
\end{abstract}
\maketitle

%% Introduction
\section{Introduction}
Sensitive measurements of temperature are essential throughout natural science and modern technology.
Increasingly detailed studies of biological, chemical, and physical processes, the miniaturisation of electronics,
and emerging quantum technology drive a need for new thermometry techniques applicable at the nanoscale and in regimes where quantum effects become important. Many new approaches are being developed \cite{Carlos2016,Mehboudi2019,PhysRevLett.103.245301,Gasparinetti2011,Neumann2013NanoLett,Lukin2013Nature,Sabin2014ScientificReports,Gasparinetti2015,Moldover2016,Palma2017ChipCooling,PhysRevApplied.10.044068,Scigliuzzo2020}, however the fundamental limits to precision thermometry are not yet fully understood. Here, we determine a tight bound on the best possible precision with which temperature can be estimated in cold quantum systems, which accounts for limitations due to imperfect measurements.

The classical picture of thermometry is that of a thermometer which is brought into thermal contact with a sample. Observing the state of the thermometer after some time conveys information about the sample temperature. A similar picture can be applied in the quantum regime, where an individual quantum probe, e.g.~a two-level system, may interact with a sample system in a thermal state, and subsequently be measured to estimate the temperature. If the probe reaches thermal equilibrium with the sample, or a non-equilibrium steady state, optimal designs of the probe and of the probe-system interaction can be determined \cite{Correa2015PRL,PhysRevA.96.062103,Mehboudi2019PRL,PhysRevLett.119.090603, Guarnieri2019,Latune2020}. Outside of the steady state regime, it was found that access to the transient probe dynamics may outperform the steady-state protocols \cite{PhysRevA.91.012331,PhysRevLett.118.130502,PhysRevA.93.053619}, that dynamical control acts as a resource \cite{Killerich2018PRA,Feyles2019PRA,Mukherjee2019}, and that thermometry can in some cases be mapped to a phase estimation problem \cite{PhysRevA.92.032112,PhysRevA.82.011611}. These findings have spurred further investigations into non-equilibrium thermometry \cite{Cavina2018PRA,PhysRevA.96.012316,Seah2019arxiv}.

Any thermometric technique will be subject to constraints due to finite measurement resolution. In the probe-sample picture, the size of the probe will limit the amount of information which can be extracted about the sample. More generally, any measurement on the sample, implemented using a finite-sized apparatus, comes with a lower bound on the attainable resolution of e.g.\ the system energy spectrum \cite{Frowis2016PRL,Potts2019fundamentallimits,Guryanova2019arxiv}. Similar restrictions apply in situations where measurements can be made on only part of a large sample \cite{Pasquale2015Nature,Hovhannisyan2018PRB,PhysRevA.95.052115}, and clearly such finite-resolution constraints must play an important role in formulating fundamental bounds on the attainable thermometric sensitivity.

Here, we derive a bound on the precision scaling with temperature, as the temperature approaches zero, for thermometers with finite energy resolution. Our bound applies to any thermometric technique based on measurements which do not resolve the individual energy levels of the sample energy spectrum. We furthermore demonstrate that this scaling can be attained using a single-qubit probe, showing that the bound is tight. To derive our bound, we build upon the framework for finite-resolution quantum thermometry introduced by Potts, Brask, and Brunner in \cite{Potts2019fundamentallimits}.

Our results also demonstrate that thermometry with a vanishing absolute error at low temperature is possible with finite resolution, answering an interesting question left open by previous work \cite{Potts2019fundamentallimits,Hovhannisyan2018PRB,Razavian2019EPJ}. For systems with a heat capacity that vanishes at low temperatures, a property often included in the third law of thermodynamics, the relative error must diverge, regardless of the available resolution \cite{Potts2019fundamentallimits}. The absolute error may either also diverge, stay constant, or vanish, with the latter thus being the best behaviour one can hope for. However, for gapped systems, even the absolute error in any unbiased temperature estimate must diverge when the temperature becomes comparable to the gap \cite{Paris2016JPAMT}. A constant or vanishing absolute error, on the other hand, has been seen in gapless systems, when employing a measurement with a continuous outcome implying an infinite resolution~\cite{Hovhannisyan2018PRB}. Our results show that a vanishing absolute error may be obtained with a finite-resolution measurement having as little as two outcomes.

This paper is organized as follows. In Sec.~\ref{sec:temperature_estimation} we introduce a general temperature estimation procedure, following \cite{Potts2019fundamentallimits}, and discuss the fundamental precision bounds imposed by the third law of thermodynamics. In Sec.~\ref{sec:scaling_bound} we propose a finite-resolution criterion, and show how this criterion leads to a tight bound on the attainable precision. In Sec.~\ref{sec:noisy_measurements} we generalize the framework to include noisy measurements, and finally in Sec.~\ref{sec:spin_boson_model} we investigate a single-qubit thermometer coupled to a bosonic bath, showing that our bound can be saturated in a physical scenario. Our analytical results are supported by numerical simulations of the temperature estimation procedure.

%%%%%%%%%%%%%%%%%%%%%%%%%%%%%%%%%%%%%%%%%%%%%%%

\section{Temperature estimation} \label{sec:temperature_estimation}
We consider a quantum system described by the canonical thermal state $\rho_{\beta}=\exp\left[-\beta H\right]/\mathcal{Z}_{\beta}$,
with $H$ the Hamiltonian operator of the system,
and $\mathcal{Z}_{\beta} \equiv \tr \left\lbrace \exp\left[-\beta H\right] \right\rbrace$ the canonical partition function.
The thermal state is parameterized by an inverse temperature $\beta = 1/k_{B}T$ where $k_{B}$ is the Boltzmann constant.
For convenience we adopt units in which $k_{B}=1$, such that temperature has the units of energy.
The task we consider is how to estimate the temperature $T$ of the system.
We remark that throughout we consider thermal states where the temperature does not itself fluctuate.
However, since temperature is not directly measurable (it is not a quantum mechanical observable),
there are fluctuations in any temperature estimate based on indirect measurements.

\subsection{Quantifying the estimation precision}
A general temperature estimation procedure consists of first performing a measurement on the system.
The most general $N$-outcome measurement is represented by a positive-operator valued measure ($\text{\small{POVM}}$) with $N$ elements $\Pi_{m}$.
Such $\text{\small{POVM}}$s capture any possible measurement in quantum mechanics,
including scenarios in which information is obtained through a probe interacting with the system,
as well as those exploiting quantum coherence \cite{Sabin2014ScientificReports,PhysRevA.91.012331,PhysRevLett.118.130502}.
Each $\text{\small{POVM}}$ element $\Pi_{m}$ corresponds to a measurement outcome $m$ which is observed with probability
\begin{equation} \label{eq:probability}
    p_{m; \beta} =  \tr \left\lbrace \Pi_{m} \rho_{\beta} \right\rbrace ,
\end{equation}
and the resulting probability distribution encodes the system temperature as a statistical parameter.
The second step in estimating the temperature is to construct an estimator $T_{\text{est}}$.
A general prescription for doing this does not exist \cite{Paris2009IJQI}.
However, it can be shown that for any unbiased estimator the variance is lower bounded through the Cramer-Rao inequality
$\delta T_{\text{est}}^{2} \geq 1/\nu \mathcal{F}_{T}$ \cite{Cramer1999},
where $\nu$ is the number of independent measurement rounds and
\begin{equation} \label{eq:fisher_info}
    \mathcal{F}_{T} \equiv \sum_{m=1}^{N} p_{m;\beta} \left[ \partial_{T} \ln p_{m;\beta} \right]^{2},
\end{equation}
is the Fisher information.
We note that the Cramer-Rao inequality is asymptotically tight for Bayesian or maximum likelihood estimators \cite{Paris2009IJQI}.
Throughout, motivated by the Cramer-Rao inequality, we adopt the Fisher information as the quantifier of precision.

Identifying measurement strategies for which the temperature estimate can achieve minimal variance
corresponds to maximizing the Fisher information over all possible measurements ($\small{\text{POVM}}$s).
This results in a measurement-independent quantity, the quantum Fisher information $\mathcal{F}_{T}^{Q}$ \cite{Braunstein1994PRL} .
Within the canonical ensemble, it can be shown that a projective measurement of the system energy is optimal \cite{Paris2016JPAMT,Potts2019fundamentallimits}.
The quantum Fisher information is then related to the variance of the system energy
\begin{equation} \label{eq:energy_variance}
    T^{4} \mathcal{F}^{Q}_{T} = \left\langle H^{2} \right\rangle - \left\langle H \right\rangle^{2},
\end{equation}
where $\left\langle O \right\rangle=\tr \left\lbrace O \rho_{\beta} \right\rbrace$.
This expression provides a fundamental upper bound on the attainable value of the Fisher information for any measurement at any temperature.
As a consequence of the third law of thermodynamics,
or more explicitly the assumption that the heat capacity vanishes at zero temperature,
the variance of the system energy must vanish at least quadratically in temperature as absolute zero is approached \cite{Potts2019fundamentallimits}.
Hence it follows that $T^{2}\mathcal{F}^{Q}_{T}$ must vanish in the low-temperature limit,
and that the relative error $\delta T_{\text{est}}^{2}/T^{2}$ must diverge by virtue of the Cramer-Rao inequality.
This relation constitutes the ultimate bound on the optimal low-temperature scaling behaviour of the Fisher information,
applicable for any system and for any measurement strategy.

\begin{figure}
    \begin{center}
	\includegraphics[width=7.5cm]{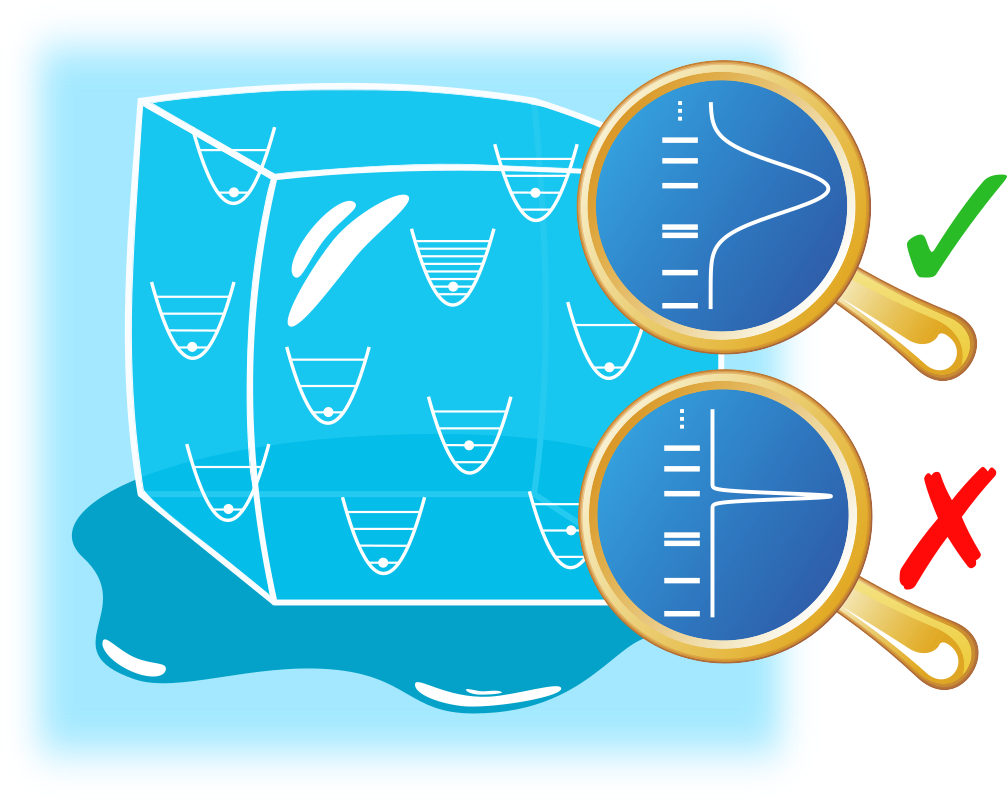}
	\caption{Finite measurement resolution is interpreted as an inability to sharply distinguish between consecutive system energy eigenstates and results in a non-trivial constraint on the attainable thermometric precision. For a macroscopic system with an effectively continuous energy spectrum, any measurement is subject to finite resolution and thus limited by the bound in Eq.~\eqref{eq:bound2}.}
    \label{fig:concept}
    \end{center}
\end{figure}

\subsection{Accounting for measurement limitations}
In many settings of interest, it is not realistic to implement a projective measurement of the system energy.
For instance, whenever the gaps in the energy spectrum are below the energy resolution of the available measurement \cite{Hovhannisyan2018PRB},
which happens, e.g.,\ when the system is large enough to appear continuous while the measurement apparatus has a finite size,
or whenever only a finite part of the full system can be interacted with within a finite time (see Fig.~\ref{fig:concept}).
Under such conditions of constrained experimental access,
it is useful to introduce the $\text{\small{POVM}}$ energies \cite{Potts2019fundamentallimits}
\begin{equation} \label{eq:spectrum}
	E_{m;\beta} \ \equiv \ \frac{1}{p_{m;\beta}}\tr \left\lbrace \Pi_{m} H \rho_{\beta} \right\rbrace ,
\end{equation}
where $E_{m;\beta}$ may be interpreted as the best guess of the system energy before the measurement,
given that outcome $m$ was observed \cite{Potts2019fundamentallimits}.
In the case of projective energy measurements on the system, the $\text{\small{POVM}}$ energies coincide with the system energy eigenvalues.
In general however, the $\text{\small{POVM}}$ energies are temperature dependent.

For convenience we may identify a specific $\text{\small{POVM}}$ energy $E_{0;\beta}$,
defined as the smallest $\text{\small{POVM}}$ energy in the low-temperature limit.
We can then introduce the $\text{\small{POVM}}$ energy gaps $\Delta_{m;\beta} \equiv E_{m;\beta} - E_{0;\beta}$, which by definition are non-negative at low temperatures.
In terms of these gaps, the Fisher information for a general measurement is given by
\begin{equation} \label{eq:povm_variance}
    \mathcal{F}_{T} \ = \ \frac{\sum_{m}p_{m;\beta}\Delta_{m;\beta}^{2} - \left( \sum_{m}p_{m;\beta}\Delta_{m;\beta}\right)^{2}}{T^{4}} .
\end{equation}
Similarly to the quantum Fisher information, the above expression takes the form of an energy variance.
However for general measurements the energy spectrum of the system is replaced by the spectrum of $\text{\small{POVM}}$ energies,
and the Boltzmann probabilities associated with projective energy measurements are replaced by the $\text{\small{POVM}}$ probabilities.
These changes incorporate restrictions due to limitations of the measurement on top of those imposed by the system itself.

In investigating the scaling behaviour we are implicitly assuming that the Fisher information is a continuous function of temperature,
which implies that the $\text{\small{POVM}}$ energy gaps $\Delta_{m;\beta}$ must also be continuous functions.
Following Ref.~\cite{Potts2019fundamentallimits},
we are going to study the scaling behaviour of the Fisher information when the $\text{\small{POVM}}$ energy gaps have a well-defined power-series expansion in temperature around absolute zero
\begin{equation} \label{eq:spectrum_expansion}
	\Delta_{m;\beta} = \Delta_{m,0} + \sum_{k=1}^{\infty} \Delta_{m,k} \beta^{-k} .
\end{equation}
By virtue of Weierstrass' approximation theorem, any continuous function can be approximated arbitrarily well by such a power series~\cite{Jeffreys2000}.
Note that this formulation does not exclude the case of projective energy measurements as this would be described by a series with only the constant term.
For more general measurements, however, the expansion might contain non-zero higher-order coefficients.

Following Potts et al.~\cite{Potts2019fundamentallimits} we can make use of the relation between the $\small{\text{POVM}}$ energies and the associated probabilities (Eq. \eqref{eq:spectrum}) to write $\Delta_{m;\beta} = - \partial_{\beta} \ln p_{m;\beta}/p_{0;\beta}$. Given the power-series expansion of the $\text{\small{POVM}}$ energy gaps, we can integrate this equation and express the ratio of the probabilities for outcomes $m$ and 0 as
\begin{equation} \label{eq:formal_distribution}
    \frac{p_{m;\beta}}{p_{0;\beta}} = g_{m} e^{-\beta \Delta_{m,0}} \beta^{-\Delta_{m,1}} \prod_{k=1}^{\infty} e^{\Delta_{m,k+1}\beta^{-k}/k} ,
\end{equation}
where $g_{m}$ is a temperature-independent integration constant. We stress that as a consequence of how we defined $E_{0;\beta}$, the probability $p_{0;\beta}$ is the largest probability at zero temperature and must be non-vanishing in this limit. We thus obtain an expression for the probability of obtaining outcome $m$ given fully in terms of the expansion coefficients of the corresponding $\small{\text{POVM}}$ energy gap (note that the explicit dependence on $p_{0;\beta}$ could be avoided by using the fact that the full distribution must be normalised).

\subsection{Low-temperature scaling behaviour}
The above model of limited measurements allows us to obtain,
by substituting Eqs.~\eqref{eq:spectrum_expansion} and ~\eqref{eq:formal_distribution} into Eq.~\eqref{eq:povm_variance},
an expression for the Fisher information given fully in terms of the $\small{\text{POVM}}$ energy gaps. Based on this, we can analyse the possible scaling behaviour of the Fisher information, as the system approaches zero temperature. First of all, we note that Eq.~\eqref{eq:povm_variance} can be rewritten as
\begin{equation} \label{eq:fisher_low_temperatures}
    \mathcal{F}_{T}
    = \frac{1}{2 T^2}\sum_{m,n} p_{m;\beta}p_{n;\beta} \left( \beta\Delta_{m;\beta} - \beta\Delta_{n;\beta} \right)^{2} .
\end{equation}
Notice that all terms on the right-hand side are positive,
and because of this the scaling behaviour of the Fisher information is determined by the term in the sum (or the set of terms) which vanishes least rapidly as the temperature goes to zero. We now consider the scaling that arises from different terms in Eq.~\eqref{eq:fisher_low_temperatures}. We focus on terms that result in sub-exponential scalings, referring the reader to Ref.~\cite{Potts2019fundamentallimits} for a discussion of the remaining terms.

For convenience, we define the ground-state set of measurement outcomes, as those for which the probability of obtaining that outcome remains finite at zero temperature (note that the outcome $m=0$ is in the ground-state set by definition). From Eq.~\eqref{eq:formal_distribution}, we see that formally this set can be defined as $\Omega = \{m \, | \, \Delta_{m,0} = \Delta_{m,1} = 0 \}$.
Now consider those terms in the Fisher information above where both outcomes belong to the ground-state set. To leading order in temperature, the contribution from these terms takes the form
\begin{equation} \label{eq:gs_contribution}
    \frac{1}{2 T^2} \sum_{m,n \in \Omega} p_{m;\beta}p_{n;\beta} \left( \Delta_{m,j} - \Delta_{n,j} \right)^{2} T^{2(j-1)} ,
\end{equation}
where $j$ labels the lowest order for which the expansion coefficient of any element in the ground-state set is non-zero ($j\geq 2$). These terms in the sum thus vanish at least quadratically, giving at best a constant contribution to the Fisher information. Notice that if the ground-state set contains only a single outcome ($m=0$), then the contribution is identically zero.

Next we consider the terms in the Fisher information where one of the outcomes belong to the ground-state set but the other one does not. To this end, we define the set of outcomes $\tilde{\Omega} = \{m \, | \, \Delta_{m,0} = 0$ and $\Delta_{m,1} \neq 0 \}$,
for which the associated probability vanish sub-exponentially as the temperature goes to zero.
The set of outcomes $\tilde{\Omega}$ has an associated $\small{\text{POVM}}$ energy coinciding with that of the ground-state set at zero-temperature, but exhibits a linear degeneracy splitting at finite temperature.
To leading order in temperature, the contribution from the corresponding terms is
\begin{equation} \label{eq:es_contribution}
    \frac{1}{T^2}\sum_{m \in \tilde{\Omega} } g_{m} \Delta_{m,1}^{2} T^{\Delta_{m,1}} ,
\end{equation}
which vanishes at a rate determined by the the first-order expansion coefficients $\Delta_{m,1}$. It is straightforward to show that all other contributions vanish exponentially in the low-temperature limit.

The (sub-exponential) low-temperature behaviour of the right-hand side of the Fisher information \eqref{eq:fisher_low_temperatures},
is generally given by the sum of Eq.~\eqref{eq:gs_contribution} and Eq.~\eqref{eq:es_contribution}.
Which of these two dominate depends on the smallest first-order expansion coefficient.
If the set $\tilde{\Omega}$ is not empty,
and at least one element in the set has a value $\Delta_{m,1}<2(j-1)$,
where $j$ denotes the lowest order with non-vanishing expansion coefficient within the ground-state set,
then the low-temperature behaviour of the Fisher information is captured by
\begin{equation} \label{eq:fisher_omega_tilde}
    \mathcal{F}_{T} = \sum_{m \in \tilde{\Omega} } g_{m} \Delta_{m,1}^{2} T^{\Delta_{m,1}-2} .
\end{equation}
In principle the first-order coefficient can take any positive value
without violating the scaling bound imposed by the third law of thermodynamics (ensuring divergence of the relative error). Notice that even a divergent low-temperature behaviour of the Fisher information can in principle be realised, if $\Delta_{m,1}$ can take a value smaller than 2.

%%%%%%%%%%%%%%%%%%%%%%%%%%%%%%%%%%%%%%%%%%%%%%%

\section{Scaling bound for large systems} \label{sec:scaling_bound}

In this section, we propose a finite-resolution criterion characterizing realistic measurements. We aim to capture any situation in which the available measurements cannot resolve the individual gaps in the system energy spectrum, which therefore appears continuous. Below, we make this statement precise. We then go on to show how this criterion leads to a lower bound on the first-order coefficient $\Delta_{m,1}$, constraining the low-temperature scaling of the error in any temperature estimation scheme. Furthermore we present an example of a measurement saturating the finite-resolution bound, showing that the bound is tight.

\subsection{Finite-resolution criterion}
In the regime where the system has an effectively continuous energy spectrum (as the measurement only resolves energy differences much larger than the gaps in the spectrum, it is convenient to work with the system density of states $\mathcal{D}(\epsilon) \equiv \sum_{k} d_{k} \delta(\epsilon-\epsilon_k)$, where the sum is over distinct system energy eigenvalues and $d_{k}$ is the corresponding degeneracy. Throughout, we adopt the convention that the smallest system energy eigenvalue is set to zero ($\epsilon_{0} = 0$).

Now, we introduce a filtered density of states $\mathcal{D}_{m}$ for each measurement outcome $m$,
as the system density of states filtered through the corresponding $\small{\text{POVM}}$ element
\begin{equation}
    \mathcal{D}_{m}(\epsilon) \equiv \sum_{k} d_{k} \delta(\epsilon-\epsilon_{k}) \tr \left[ \Pi_{m} \frac{\Id_{\epsilon_{k}}}{d_{k}} \right] ,
\end{equation}
where $\Id_{\epsilon_{k}}$ is the projection operator onto the eigenspace with energy $\epsilon_{k}$.
Notice that the sum of all the filtered densities of states adds up to the total density of states.
Furthermore, we introduce the continuous filter function $f_{m}(\epsilon)$,
formally defined by the values $f_{m}(\epsilon_{k})= \tr \left[ \Pi_{m} \Id_{\epsilon_{k}}/d_{k} \right]$ and the straight-line segments connecting these values.
In addition we note that the density of states can be expressed as the rate of change of the number of states with energy below $\epsilon$
$\sigma(\epsilon) = \sum_{k} d_{k} \theta(\epsilon-\epsilon_{k})$, where $\theta$ denotes the Heaviside step function.
Given these, the filtered density of states decompose into the product
\begin{equation}
    \mathcal{D}_{m}(\epsilon) = f_{m}(\epsilon) \frac{d\sigma(\epsilon)}{d\epsilon} ,
\end{equation}
where the filter function fully characterizes the implemented measurement.
Importantly we notice that the function $\sigma(\epsilon)$ is non-decreasing for all energies.
If we compute the Laplace transform in $\beta$ of the filtered density of states,
the result takes the form of a Stieltjes integral over a measure given by $\sigma(\epsilon)$ \cite{Ballentine2014}
\begin{equation}
\label{eq:dmproba}
    \hat{\mathcal{D}}_{m}(\beta) \equiv \int_{0}^{\infty} d\sigma(\epsilon) f_{m}(\epsilon) e^{-\beta \epsilon} = \mathcal{Z}_{\beta}p_{m;\beta} .
\end{equation}
The last equality can be obtained directly from equation~\eqref{eq:probability},
and relates the Laplace transformed filtered density of states to the product of the probability and the canonical partition function.
Notice that the measure $\sigma(\epsilon)$ is a discontinuous function of energy.

For macroscopic systems the measure can often be approximated by an effective continuous measure,
when $\sigma(\epsilon)$ and $f_{m}(\epsilon)$ vary on widely separated energy scales.
To see this, we first define the averaged measure with respect to an energy window $\omega$ by
\begin{equation}
    \sigma_{\omega}(\epsilon) = \theta(\epsilon) \frac{1}{\omega} \int_{\epsilon-\omega/2}^{\epsilon+\omega/2} ds \sigma(s) ,
\end{equation}
which for non-zero $\omega$ is a continuous function of energy except at $\epsilon = 0$,
and which tends to a differentiable function of energy as $\omega$ is increased.
The inclusion of the step function at zero energy is important if we are to capture the zero temperature limit correctly,
since it ensures that the ground-state of the averaged model coincides with that of the exact model.
For the purposes of low-temperature thermometry only the low-energy behaviour is of importance,
and to leading order in energy we adopt an effective measure given by
\begin{equation}
    d\sigma_{\omega}(\epsilon) = \left[ d_{0;\omega} \delta(\epsilon) + \alpha_{\omega} \gamma_{\omega} \epsilon^{\gamma_{\omega}-1}  + \mathcal{O}(\epsilon^{\gamma_{\omega}}) \right] d\epsilon,
\end{equation}
where $d_{0;\omega}$ is an effective ground-state degeneracy and $\alpha_{\omega}$,$\gamma_{\omega}$ are positive, real-valued constants.
The coefficient $\gamma_{\omega}>0$ characterizes the low-energy growth in the total number of states with energy less than $\epsilon$.

If we compute the Laplace transform with respect to this averaged measure (which now takes the form of a standard Riemann integral) we obtain to leading order in energy
\begin{equation} \label{eq:continous_Dm}
\begin{aligned}
    \hat{\mathcal{D}}_{m;\omega}(\beta) = \
    & d_{0;\omega}f_{m}(0) \\
    & + \alpha_{\omega} \gamma_{\omega} \int_{0}^{\infty} d\epsilon \epsilon^{\gamma_{\omega}-1} f_{m}(\epsilon) e^{-\beta \epsilon} .
\end{aligned}
\end{equation}
The averaged measure tends to overestimate the number of low-energy states as $\omega$ is increased,
however this effect becomes negligible in the limit $\omega\ll T$.
Now if we assume that $f_{m}(\epsilon)$ does not vary significantly across an energy range $\omega$,
then $\hat{\mathcal{D}}_{m}(\beta)$ is well approximated by the averaged function $\hat{\mathcal{D}}_{m;\omega}(\beta)$.
More quantitatively we can state this condition in the form of an inequality
\begin{equation} \label{eq:discrete_criterion}
    \frac{ \mid  f_{m}(\epsilon+\omega) - f_{m}(\epsilon)\mid  }{\omega} \ll \frac{1}{\omega} ,
\end{equation}
which bounds the rate of change of the filter function with energy.
For macroscopic systems we can take the limit $\omega \rightarrow 0$,
and in this case we are going to adopt the following \textit{finite-resolution criterion (FRC)}: \\

\begin{quote}
\textbf{FRC:} \textit{In the limit of a macroscopic system, the filter function $f_{m}(\epsilon)$ tends to a continuous, right-differentiable function of the system energy.}
\end{quote}
This is nothing more than a restatement of equation~\eqref{eq:discrete_criterion} for vanishing $\omega$,
which restricts the rate of change of the filter function to a finite value. We note that at $\epsilon=0$, the filter function may be discontinuous and Eq.~\eqref{eq:discrete_criterion} tends to the right derivative for $\omega\rightarrow0$.

\subsection{Finite-resolution bound}
Having characterized what we mean by a finite-resolution measurement,
we ask what the consequences of our finite-resolution criterion are for the behaviour of the $\small{\text{POVM}}$ energy gaps in the macroscopic limit. 
By making use of equation~\eqref{eq:formal_distribution}, we obtain the relation (we now drop the dependence on the energy window $\omega$ and write simply $d_{0}$,$\alpha$ and $\gamma$)
\begin{equation} \label{eq:povm_dos_laplace}
    \hat{\mathcal{D}}_{m}(\beta) = \hat{G}_{m}(\beta) \hat{\mathcal{D}}_{0}(\beta) ,
\end{equation} 
where for convenience we have defined the transfer function
\begin{equation}
    \hat{G}_{m}(\beta) \equiv g_{m} e^{-\beta \Delta_{m,0}} \beta^{-\Delta_{m,1}} \prod_{k=1}^{\infty} e^{\Delta_{m,k+1}\beta^{-k}/k} .
\end{equation}
Now this is a relation at the level of the Laplace-transformed, filtered densities of states.
We can  obtain a relationship directly between the filtered densities of states
by taking the inverse Laplace transform of both sides of Eq.~\eqref{eq:povm_dos_laplace}.
By applying the Laplace convolution theorem \cite{Arfken2013,Cahill2013}, we derive the relation
\begin{equation} \label{eq:relation_Dm}
    \mathcal{D}_{m}(\epsilon)
    = \int_{0}^{\epsilon}ds \ G_{m}(\epsilon-s) \mathcal{D}_{0}(s) .
\end{equation}
We now focus on the specific case of $m \in \tilde{\Omega}$.
For these outcomes, the inverse Laplace transform can be computed straightforwardly, and to leading order in energy we obtain
\begin{equation} \label{eq:Gm_defapp}
G_{m}(\epsilon) = \frac{g_m}{\Gamma(\Delta_{m,1})} \epsilon^{\Delta_{m,1}-1} + \mathcal{O}(\epsilon^{\Delta_{m,1}}), 
\end{equation}
where $\Gamma(\Delta_{m,1})$ denotes the Gamma function \cite{Arfken2013}.
As we saw in the preceding section,
the outcomes within $\tilde{\Omega}$ are exactly the ones with potential to provide optimal low-temperature scaling of the Fisher information.

Recall, that the reference outcome $m=0$, was chosen such that the associated probability approaches a constant value at zero temperature.
This implies that the overlap of the $\small{\text{POVM}}$ element $\Pi_{0}$ with the system ground state is non-zero, and therefore $f_{0}(0)$ is non-zero.
On the other hand for outcomes $m \in \tilde{\Omega}$ the probability vanishes in the low-temperature limit, implying a vanishing overlap $f_{m}(0)=0$.
Hence in this case we find from equations~\eqref{eq:continous_Dm} and \eqref{eq:relation_Dm} that to leading order in energy
\begin{equation} \label{eq:filter_low_energy}
    f_{m}(\epsilon) = \frac{g_{m}d_{0}f_{0}(0)}{\alpha \gamma \Gamma(\Delta_{m,1})} \ \epsilon^{\Delta_{m,1}-\gamma} + \mathcal{O}(\epsilon^{\Delta_{m,1}+1-\gamma}) .
\end{equation}
Based on this expression we can infer constraints on the linear coefficient.
First, the requirement that $f_{m}(0) = 0$ gives the weakest constraint $\Delta_{m,1}>\gamma$. This simply expresses the fact that the Fisher information is upper bounded by the the quantum Fisher information, which scales as $T^{\gamma -2}$ for a density of states scaling as $\epsilon^{\gamma -1}$.
Further, the finite-resolution criterion restricts the rate of change to be bounded, $\frac{d}{d\epsilon}f_m(\epsilon) < \infty$. This implies a tightened scaling bound
\begin{equation}
    \Delta_{m,1}\geq 1+\gamma, \ \ \ \text{for} \ \ m \in \tilde{\Omega}.
\end{equation}
Since $\gamma>0$ by definition, this implies that the Fisher information must grow slower than $1/T$, i.e.,
\begin{equation}
    \label{eq:bound}
    \lim_{T\rightarrow 0}T\mathcal{F}_T =  0.
\end{equation}
Further note that a diverging Fisher information in the low-temperature limit can only be realized through a
$\sigma(\epsilon)$ that grows sub-linearly with energy, i.e., $\gamma<1$.
As an example of a system exhibiting such a sub-linear growth we mention systems of massive non-interacting particles at zero chemical potential \cite{Potts2019fundamentallimits}. For such systems $\gamma = 1/2$ for one-dimensional geometries. 

By virtue of the Cramer-Rao bound, Eq.~\eqref{eq:bound} implies that the absolute error (squared) must vanish more slowly than $T$
\begin{equation}
\label{eq:bound2}
\lim_{T\rightarrow 0} \frac{\delta T^2_{\rm est}}{T} = \infty.
\end{equation}
The equivalent Eqs.~\eqref{eq:bound} and \eqref{eq:bound2} constitute the main result of our paper. They imply that for an effectively continuous spectrum, the low-temperature scaling of the precision is not only bounded by the third law, which demands a diverging relative error, but by a tighter bound. Interestingly, our bound still allows for a vanishing absolute error, a scenario that can be physically realized as illustrated below.

\subsection{Proving tightness of bound} \label{sec:exp-res-meas}
We now illustrate that the proposed finite-resolution bound is tight.
Consider a binary measurement which resolves the system ground state exponentially well in the sense that it has $\small{\text{POVM}}$ elements
\begin{equation}
    \Pi_{0} = e^{-\kappa {H}} , \ \ \ \ \Pi_{1} = \Id - e^{-\kappa {H}} ,
\end{equation}
where $\kappa > 0$. Note that the overlap of $\Pi_0$ with the system energy eigenstates decays exponentially away from zero. This feature makes is straightforward to write down the filtered density of states.
Focusing on $m=1$ we find
\begin{equation}
    \mathcal{D}_{1}(\epsilon) = \left[1-e^{-\kappa \epsilon} \right] \mathcal{D}(\epsilon) ,
\end{equation}
where nothing has been assumed about the form of the system density of states.
We thus see that the corresponding filter function takes the form $f_{1}(\epsilon) = \kappa \epsilon + \mathcal{O}(\epsilon^{2})$
to leading order in energy. If we adopt the density of states introduced in the preceding subsection, that is 
$\mathcal{D}(\epsilon) = d_{0}\delta(\epsilon) + \alpha \gamma \epsilon^{\gamma - 1} + \mathcal{O}(\epsilon^{\gamma})$,
then upon comparison with equation~\eqref{eq:filter_low_energy} we find $\Delta_{1,1}=1+\gamma$.
Hence the binary exponential resolution measurement saturates the finite-resolution bound.

For good measure we now show how the same conclusion can be derived directly from the probabilities.
The probability of obtaining outcome $m=0$ can be written in terms of the system partition function as
\begin{equation}
    p_{0;\beta} = \mathcal{Z}_{\kappa + \beta} \mathcal{Z}_{\beta}^{-1} .
\end{equation}
Substituting the probabilities $p_{0;\beta}$ and $p_{1;\beta}=1-p_{0;\beta}$ into the general form of the Fisher information (Eq.~\ref{eq:fisher_info}),
one finds that
\begin{equation}
    T^{4}\mathcal{F}_{T} = \frac{\mathcal{Z}_{\kappa+\beta}}{\mathcal{Z}_{\beta}-Z_{\kappa+\beta}} 
    \left( \left\langle H \right\rangle_{\beta} - \left\langle H \right\rangle_{\kappa+\beta} \right)^{2} .
\end{equation}
The partition function is given by the Laplace transform of the density of states,
hence we find $\mathcal{Z}_{\beta} = d_{0} \exp \left( \alpha \gamma \Gamma(\gamma) \beta^{-\gamma}/d_{0} \right)$
(in \appref{app:bosonDOS} we show how this form of the partition function describes a system of non-interacting bosonic modes).
From this form of the partition function we can derive the
low-temperature behaviour of the average energy 
\begin{equation}
    \left\langle H \right\rangle_{\beta} = \frac{\alpha \gamma^{2} \Gamma(\gamma)}{d_{0}} \beta^{-(1+\gamma)} .
\end{equation}
If we substitute these into the above Fisher information, we find that to leading order in temperature (assuming that $\kappa/\beta \ll 1$)
\begin{equation}
\label{eq:fisher_exp}
    \mathcal{F}_{T} = \alpha \kappa \gamma^{2} \Gamma(\gamma) (1+\gamma)^{2} T^{\gamma-1} + \mathcal{O}(T^{2\gamma}),
\end{equation}
which takes the form of Eq.~\eqref{eq:fisher_omega_tilde} with $\Delta_{1,1} = 1+\gamma$ and $g_{1} = \alpha \kappa \gamma^{2} \Gamma(\gamma)$.
Since $\gamma$ can in principle take any positive value, the exponential-resolution measurement saturates the finite-resolution bound and asymptotically attains a Fisher information scaling as $1/T$ in the limit $\gamma\rightarrow 0$.

%%%%%%%%%%%%%%%%%%%%%%%%%%%%%%%%%%%%%%%%%%%%%%%

\section{Generalization to noisy measurements} \label{sec:noisy_measurements}
In this section we extend the thermometry framework above to include noisy measurements. As the framework is general, one might ask if noise effects are not already accounted for. The answer is that in principle noise effects are described. However, for some noisy measurements, the $\small{\text{POVM}}$ energy gap does not have a Taylor expansion. While one may still approximate the energy gap by a polynomial, a physically appealing extension of the formalism allows for circumventing this approximation. We find that our bound given in Eq.~\eqref{eq:bound2} also holds for noisy measurements.

\subsection{Noisy temperature measurements}
To model noisy measurements, we consider the case where the observed outcomes $m$ correspond to coarse graining over a fine-grained $\small{\text{POVM}}$ with elements $\Pi_{m\mu}$.
The probability of observing $m$ is then
\begin{equation}
    p_{m;\beta} = \sum_{\mu} p_{m\mu;\beta} = \sum_{\mu} \tr \left\lbrace \Pi_{m\mu} \rho_{\beta} \right\rbrace .
\end{equation}
Physically this could correspond to a measurement implemented using a sensor, where only a subset of the sensor degrees of freedom (or a subspace of the full sensor Hilbert space) is experimentally accessible. If we were to compute the Fisher information directly using the fine-grained distribution $p_{m\mu}$, we recover the noiseless results, and obtain an upper bound on the Fisher information computed from the coarse-grained distribution. This fact follows directly from the relation between the relative entropy of two probability distributions differing by an infinitesimal temperature $\delta T$ and the Fisher information
\begin{equation}
\begin{aligned}
    D\left( p_{T} \vert \vert p_{T+\delta T}\right) 
    = \mathcal{F}_{T} \delta T^{2} \ \ \ \text{as} \ \ \ \delta T \rightarrow 0 .
\end{aligned}
\end{equation}
Since the relative entropy is monotonically decreasing under coarse-graining \cite{Wilde2017},
we conclude that noise always reduces the Fisher information.

The question we now address is, how it impacts the attainable scaling with temperature.
Following the approach developed above, we introduce the fine-grained $\small{\text{POVM}}$ energies
\begin{equation} \label{eq:fine_grained_povm_energy}
    E_{m\mu;\beta} \equiv \frac{1}{p_{m\mu;\beta}} \tr \left\lbrace \Pi_{m\mu} \rho_{\beta} \right\rbrace ,
\end{equation}
which may be interpreted as the best guess of the system energy before the measurement, given the outcome $(m,\mu)$ \cite{Potts2019fundamentallimits}. For convenience we identify the smallest $\small{\text{POVM}}$ energy in the low-temperature limit with the outcome $E_{00;\beta}$, and then define the fine-grained $\small{\text{POVM}}$ energy gap $\Delta_{m\mu;\beta} \equiv E_{m\mu;\beta} - E_{00;\beta}$,
which by definition is non-negative at low temperatures. Modelling the fine-grained $\small{\text{POVM}}$ energy gaps by a power-series expansion around zero temperature as in Eq.~\eqref{eq:spectrum_expansion}, we are led to a probability distribution identical to \eqref{eq:formal_distribution},
but with $m$ replaced by the compound index $m\mu$.

Since the Fisher information is not defined with respect to the fine-grained probabilities, but rather with respect to the coarse-grained probabilities, the relevant energies are the coarse-grained $\small{\text{POVM}}$ energy gaps defined by
\begin{equation}
\label{eq:gap_cg}
    \Delta_{m;\beta}^{(c)} \equiv \sum_{\mu} \frac{p_{m\mu;\beta}}{p_{m;\beta}} \Delta_{m\mu;\beta} .
\end{equation}
In terms of these, the Fisher information can be written in the same form as the fine-grained Fisher information of Eq.~\eqref{eq:fisher_low_temperatures},
but with the fine-grained probability and the fine-grained $\small{\text{POVM}}$ energy gaps replaced by their coarse-grained versions
\begin{equation} \label{eq:coarse_fisher_info}
    \mathcal{F}_{T} = \frac{1}{2 T^2} \sum_{m,n} p_{m;\beta}p_{n;\beta} \left( \beta \Delta^{(c)}_{m;\beta} - \beta \Delta^{(c)}_{n;\beta} \right)^{2} .
\end{equation}
Notice that all terms in the sum are positive. Hence, the scaling behaviour of the Fisher information is determined by the term (or set of terms) which vanishes least rapidly as the temperature approaches zero.

From Eq. \eqref{eq:gap_cg}, we can anticipate that fine-grained energy gaps that have a Taylor expansion may result in coarse-grained gaps that do not.
This may result in qualitatively different behaviour of the fine- and coarse-grained Fisher information.
In particular, noise may render the scaling of the Fisher information worse.
In appendix \ref{app:noise} we discuss in general terms how noise impacts the attainable Fisher information scaling. In particular, we show that the noise can never result in a better scaling for the Fisher information, implying that the bound given in Eq.~\eqref{eq:bound2} also holds for noisy measurements. Here we illustrate the effect of noise with an example.

\begin{figure}
    \begin{center}
	\includegraphics[width=8.5cm]{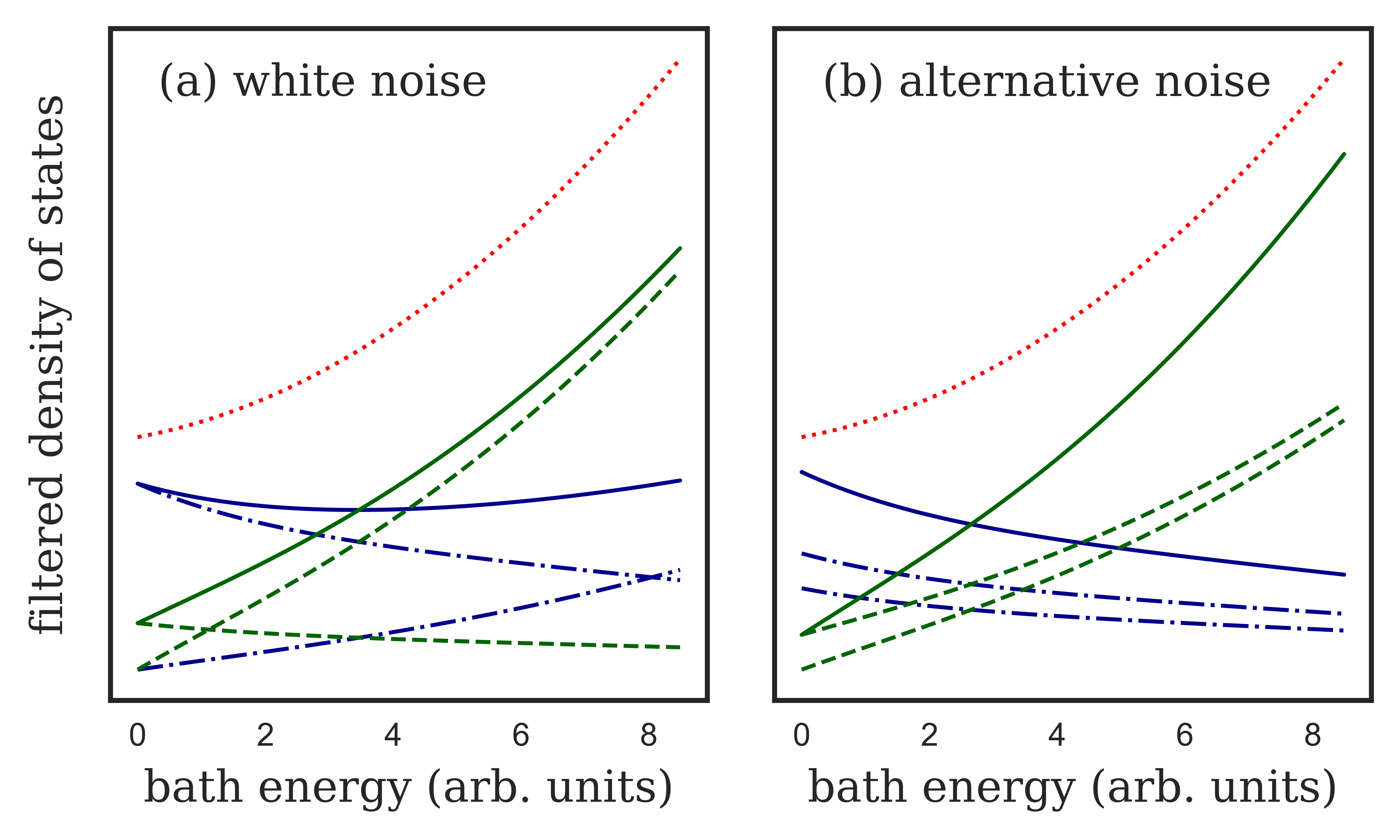}
	\caption{Illustration of filtered density of state for a noisy binary exponential resolution measurement using
	$\mathcal{D}(\epsilon) = \mathcal{L}^{-1}\left[ \exp \left( \alpha \beta^{-1} \right)\right]$ (dotted red line) with $\alpha=0.2$.
	(a) The white noise measurement corresponds to swapping the observed measurement outcomes with some probability,
	such that each coarse-grained outcome has contributions both from elements within and elements not within the ground-state set.
	The dashed green lines gives $\mathcal{D}_{00}$ and $\mathcal{D}_{01}$ (their sum is shown with the solid green line),
	and the blue dashed-dotted lines correspond to elements $\mathcal{D}_{10}$ and $\mathcal{D}_{11}$ (with their sum given by the solid blue line).
	(b) In \appref{app:noise} we show that an alternative noise model consists of a mixing of several similar measurement outcomes.
	In the specific case depicted here, the fine-grained outcomes to be summed are almost identical except for projecting onto slightly different energy distributions.}
    \label{fig:povm_dos}
    \end{center}
\end{figure}

\subsection{Illustration of noisy measurement}
\label{sec:noisy_measurements_example}
A simple example illustrating noise is obtained by adding white noise to the binary, exponential-resolution measurement of \secref{sec:exp-res-meas}.
That is, we study a binary $\small{\text{POVM}}$ defined by $\Pi_{0} = \eta \exp\left(-\kappa H\right) + (1-\eta)\Id/2$.
To understand how this noise model arises from coarse graining a fine-grained measurement,
we consider the fine-grained $\small{\text{POVM}}$
\begin{equation}
\label{eq:exppovmfine}
    \begin{aligned}
        &\Pi_{00} = \frac{1+\eta}{2} e^{-\kappa H} , \ \ \ \ \Pi_{01} = \frac{1-\eta}{2} \left( \Id - e^{-\kappa H}\right), \\
        &\Pi_{10} = \frac{1+\eta}{2} \left( \Id - e^{-\kappa H}\right) , \ \ \ \ \Pi_{11} = \frac{1-\eta}{2}e^{-\kappa H},
    \end{aligned}
\end{equation}
such that $\Pi_{0}=\Pi_{00}+\Pi_{01}$ and $\Pi_{1}=\Pi_{10}+\Pi_{11}$.
As in the noiseless case, we suppose that the average energy exhibits a power-law behaviour $\left\langle H \right\rangle_{\beta} = \alpha \beta^{-(1+\gamma)}$ at low temperatures in the macroscopic limit,
with $\alpha$ and $\gamma$ both positive. The corresponding partition function (at low temperatures) is then $\mathcal{Z}_{\beta} = \exp \left( \alpha \beta^{-\gamma}/\gamma \right)$.
For the fine-grained measurement outcomes, we find that to leading order in temperature (assuming that $\kappa/\beta \ll 1$ and $\eta<1$),
the $\small{\text{POVM}}$ energy gaps with respect to the reference $E_{00;\beta}$, take the form
\begin{equation}
    \begin{aligned}
        & \Delta_{00;\beta} = \Delta_{11;\beta} = 0 , \\
        & \Delta_{10;\beta} = \Delta_{01;\beta} = (1+\gamma)T + \mathcal{O}(T^{2+\gamma}) .
    \end{aligned}
\end{equation}
We see that the fine-grained measurement outcomes have an associated set of $\small{\text{POVM}}$ energy gaps that have a Taylor series in the low-temperature limit.
Furthermore, they exhibit a linear degeneracy splitting.
It then follows from Eq.~\eqref{eq:fisher_omega_tilde} that the Fisher information takes the form
\begin{equation}
    \mathcal{F}_{T} = \alpha \kappa (1+\gamma)^{2} T^{\gamma-1} + \mathcal{O}(T^{2\gamma}) ,
\end{equation}
which is equivalent to the noiseless form found above [cf.~Eq.~\eqref{eq:fisher_exp}].
Notice that when having access to the fine-grained distribution, both the $\small{\text{POVM}}$ energies and the resulting Fisher information is independent of the parameter $\eta$ quantifying the amount of white noise.

The picture changes when considering the coarse-grained energy gap (Eq.~\eqref{eq:gap_cg}).
To leading order in temperature this is given by
\begin{equation}
    \Delta_{1;\beta}^{(c)} = \frac{1+\eta}{1-\eta} \alpha \kappa (1+\gamma) T^{2+\gamma} + \mathcal{O}(T^{3+2\gamma}).
\end{equation}
Notice that in contrast to the fine-grained energy gaps, this coarse-grained gap does not have a Taylor expansion.
Computing the Fisher information over the coarse-grained gaps and probabilities (making use of Eq.~\eqref{eq:coarse_fisher_info}) gives
\begin{equation}
    \mathcal{F}_{T} = \frac{4\eta^{2}}{1-\eta^{2}}\left(\alpha \kappa(1+\gamma)\right)^{2} T^{2\gamma} + \mathcal{O}(T^{1+3\gamma}).
\end{equation}
This example thus illustrates how noise can result in a coarse-grained gap that has no Taylor expansion and how this may result in a different (worse) scaling for the Fisher information at low-temperatures.
Qualitatively we can understand the altered scaling by studying the coarse-grained filtered density of states.
For the example considered here we have
\begin{equation}
\begin{aligned}
    & \mathcal{D}_{00}(\epsilon)= f_{00}(\epsilon)\mathcal{D}(\epsilon) = \frac{1+\eta}{2} e^{- \kappa \epsilon} \mathcal{D}(\epsilon) , \\
    & \mathcal{D}_{01}(\epsilon)= f_{01}(\epsilon)\mathcal{D}(\epsilon) = \frac{1-\eta}{2} \left( 1-  e^{- \kappa \epsilon}\right) \mathcal{D}(\epsilon) ,
\end{aligned}
\end{equation}
and under coarse-graining these are added together.
Notice that whereas the filter function $f_{01}(\epsilon)$ goes to zero as $\epsilon \rightarrow 0$, this is not true of $f_{00}(\epsilon) + f_{01}(\epsilon)$ (the same feature is found for the $m=1$ outcomes).
Hence in this case the noise removes outcomes from the set $\tilde{\Omega}$, resulting in the worse scaling (note that a vanishing filter function at $\epsilon=0$ implies a vanishing probability at $T=0$ and vice versa, cf.~Eq.~\eqref{eq:dmproba}). This effect is illustrated in Fig.\ref{fig:povm_dos}a. In \appref{app:noise} we study an alternative noise model. In this model each coarse-grained outcome can be seen as the sum of several similar (in the sense of preparing similar energy distributions) fine-grained outcomes. This is illustrated in Fig.\ref{fig:povm_dos}b.

The noisy framework put forward here shows that our finite-resolution bound, as well as the results of Ref.~\cite{Potts2019fundamentallimits} apply for any $\small{\text{POVM}}$ that can be written as a coarse graining over a fine-grained $\small{\text{POVM}}$ which has a spectrum with a well defined Taylor series. As the coarse-grained $\small{\text{POVM}}$ itself may not have a spectrum with a well defined Taylor series, this extends the applicability of the results of Ref.~\cite{Potts2019fundamentallimits} (as long as we do not want to rely on approximate Taylor series in the spirit of the Weierstrass theorem).

%%%%%%%%%%%%%%%%%%%%%%%%%%%%%%%%%%%%%%%%%%%%%%%

\section{Single-qubit probe} \label{sec:spin_boson_model}
\subsection{Measurement protocol}
We now illustrate our results by considering temperature estimation of a system of non-interacting bosons using a single qubit as a probe. The system is described by a spectrum of single-particle energies $\omega_{k}$ (we take $\hbar = 1$). Consider the following measurement strategy: (i) first we initialise the probe qubit in its ground state $\ket{0}$, (ii) then an interaction is turned on between the probe and the system for a short time $t$,
and (iii) we perform a projective measurement of the qubit energy.
Given this protocol, the probability of finding the qubit in the excited state $\ket{1}$ is
\begin{equation}
\label{eq:qubit_p1}
    p_{1;\beta} = \tr \left\lbrace \bra{0}U^{\dagger}_{t}\ket{1} \! \bra{1}U_{t}\ket{0} \rho_{\beta}\right\rbrace .
\end{equation}
We take the time-evolution operator $U_{t}$ to be generated by a time-independent Hamiltonian
\begin{equation}
    H = \sum_{k} \omega_{k} a_{k}^{\dagger}a_{k} + \frac{\Omega}{2} \sigma_{z} + H_{\text{int}} ,
\end{equation}
where $a_{k}^{\dagger},a_{k}$ denotes the bosonic creation and annihilation operators.
The probe qubit is characterised by the three Pauli operators $\left\lbrace \sigma_{x},\sigma_{y},\sigma_{z} \right\rbrace$,
and we take the probe energy to be proportional to the $\sigma_{z}$ operator.

Computing the outcome probabilities requires specifying an interaction Hamiltonian and determining the resulting dynamics.
This task is complicated by the fact that the low-temperature and short-time regime is generally not accessible via standard Markovian master equations \cite{Vega2017RMP,Mehboudi2019PRL}. However, if the interaction time is sufficiently short we can make analytical progress by approximating the probability up to second order in $t$.
In this case we find that
\begin{equation} \label{eq:prob_approx}
    p_{1;\beta} = t^{2} \tr \left\lbrace \bra{0}H_{\text{int}}\ket{1} \! \bra{1}H_{\text{int}}\ket{0} \rho_{\beta} \right\rbrace 
    + \mathcal{O}(t^{4}).
\end{equation}
We consider a linear interaction Hamiltonian consisting of an excitation-preserving part and a non-excitation-preserving part.
Introducing the raising and lowering operators $\sigma_\pm = \frac{1}{2}(\sigma_x \pm i \sigma_y)$ for the probe qubit,
the interaction Hamiltonian takes the form
\begin{equation}
    \begin{aligned}
         H_{\text{int}} = 
         & \sum_{k} g_{k} \left[ \sigma_+ a_{k} + \sigma_- a_{k}^{\dagger} \right] \\
         & + \sum_{k} \lambda_{k} \left[ \sigma_- a_{k} + \sigma_+ a_{k}^{\dagger} \right] ,
    \end{aligned}
\end{equation}
where $\left\lbrace g_{k}, \lambda_{k} \right\rbrace$ are real-valued coupling coefficients. In the limit of a macroscopic system, these coupling coefficients are taken to approach continuous functions.
Physically this means that the interaction cannot selectively probe an individual system mode (ensuring that the finite resolution criterion is satisfied).

Given $H_{\text{int}}$, it becomes straightforward to show from Eq.~\eqref{eq:prob_approx} that the excited-state probability at short times takes the form
\begin{equation}
\label{eq:pexcited}
\begin{aligned}
    p_{1;\beta} = \
    t^{2} \sum_{k} \left( g_{k}^{2} + \lambda_{k}^{2} \right) n_{\beta}(\omega_{k})
    + t^{2} \sum_{k} \lambda_{k}^{2},
\end{aligned}
\end{equation}
where $n_{\beta}(\omega_{k})$ denotes the Bose-Einstein distribution.
We see that the probability consists of two contributions: a temperature-dependent term, in which the probability is directly related to the occupation of the bath modes,
and a temperature-independent term.
The presence of the temperature-independent term means that the probability of finding the probe qubit in the excited state is generally non-zero even at arbitrarily low temperatures. As in the example in Sec.~\ref{sec:noisy_measurements_example}, this prevents a scaling of the form of Eq.~\eqref{eq:fisher_omega_tilde} and can be captured by our framework for noisy thermometry.

\subsection{Excitation-preserving interaction}
We now focus on the excitation-preserving case ($\lambda_{k}=0$), and consider an interaction characterised by a continuous spectral density of the form 
\begin{equation}
\label{eq:spectraldens}
    \rho(\omega) = \sum_{k} g_{k}^{2} \delta(\omega-\omega_{k}) = 2 \alpha \omega_{c}^{1-s}\omega^{s} e^{-\omega/\omega_{c}} ,
\end{equation}
where $\alpha$ is the dissipation strength, $s$ is the ohmicity and $\omega_{c}$ is the cutoff energy \cite{BreuerPetruccione2002,Carmichael2003,Weiss2012,Vega2017RMP}.
The sum in the excited-state probability \eqref{eq:pexcited} is then replaced by an integral, which can be solved analytically. In the low-temperature limit we find
\begin{equation}
    p_{1;\beta} = 2 \alpha \left( \omega_{c} t\right)^{2} \Gamma(1+s) \left(\frac{T}{\omega_{c}}\right)^{1+s} + \mathcal{O}(T^{2+s}).
\end{equation}
We see that this protocol gives a probability vanishing sub-exponentially as the temperature goes to zero, and comparing with the general expression Eq.~\eqref{eq:formal_distribution}, we see that to lowest order, the $\small{\text{POVM}}$ gap scales as $\Delta_{1}=\left(1+s\right)T$. The case of an excitation-preserving interaction can thus (for short time at least) be described within our noiseless thermometry framework.

From the value of the linear expansion coefficient, $\Delta_{1,1} = 1 + s$, it follows that for ohmicity approaching zero, the finite-resolution bound $\Delta_{1,1} \geq 1$ is approached. The corresponding Fisher information scales as $\mathcal{F}_{T} \propto T^{s-1}$ and thus diverges for sub-Ohmic baths in the low-temperature limit. This serves as an illustration that the finite-resolution bound is in principle attainable via an excitation-preserving interaction in the short-time limit, and thus the bound is tight. Realising such an excitation-preserving interaction may however be challenging.

\begin{figure}
    \begin{center}
	\includegraphics[width=8.0cm]{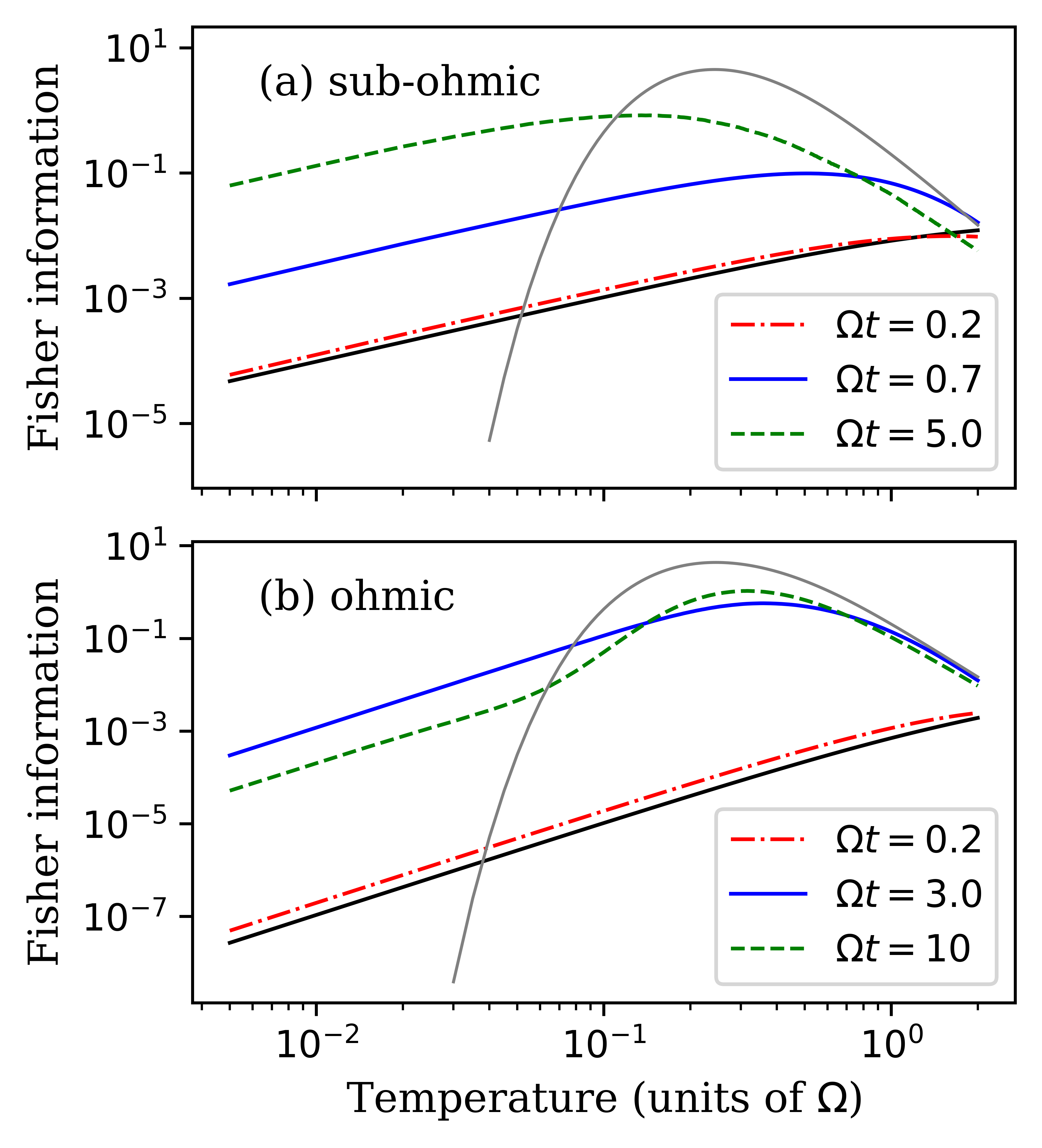}
	\caption{Numerically computed Fisher information for (a) the sub-Ohmic $(s=1/2)$, and (b) the Ohmic $(s=1)$ spin-boson model, with $\delta t = 0.1/\Omega$, $\alpha=0.1$ and $\omega_{c} = 10 \Omega$. The solid black lines display the short-time analytical results at time $\Omega t = 0.2$, showing good agreement with the numerical simulations. In case (b) the simulations exhibit a quadratic temperature scaling at low temperatures, while in case (a) a linear scaling is obtained. The solid grey line gives the Fisher information obtained from the steady state of a secular Born-Markov master equation, which scales exponentially at low temperatures \cite{Potts2019fundamentallimits}.}
    \label{fig:FI_sbm}
    \end{center}
\end{figure}

\subsection{Excitation-non-preserving interaction}
We now turn to the arguably more realistic excitation non-preserving case. The case $\lambda_{k}=g_{k}$ corresponds to the well-known spin-boson model \cite{PhysRevLett.97.016802,BreuerPetruccione2002,Weiss2012,Carmichael2003}. Adopting the same spectral density as above, the excited-state probability in this case takes the form
\begin{equation}
\label{eq:p1nonpres}
\begin{aligned}
    p_{1;\beta} = \
    & 4 \alpha \left( \omega_{c} t\right)^{2} \Gamma(1+s) \left(\frac{T}{\omega_{c}}\right)^{1+s} \\
    & + 2 \alpha \left( \omega_{c} t\right)^{2} \Gamma(1+s) + \mathcal{O}(T^{2+s}) .
\end{aligned}
\end{equation}
In contrast to the excitation-preserving case, this probability does not in general correspond to the noiseless version of Eq.~\eqref{eq:formal_distribution} since the $\small{\text{POVM}}$ energy gap $\Delta_1\propto T^{s+2}$,
does not have a Taylor expansion for arbitrary $s$ at low temperatures.
However, as shown in \appref{app:noisy_sbm}, this scenario can be described using a fine-grained POVM with energy gaps that do have a Taylor expansion.
Therefore, the scenario is captured by the noisy framework.

Given the probability \eqref{eq:p1nonpres}, a short calculation shows that the Fisher information has a low-temperature scaling given by $\mathcal{F}_{T} \propto T^{2s}$. Again, this is in full agreement with the general noisy theory developed above. Within the spin-boson model, the Fisher information thus vanishes quadratically for an Ohmic spectral density with $s=1$, and linearly for a sub-Ohmic spectral density with $s=1/2$.

To corroborate the analytical results based on the short-time approximation, we turn to a numerical simulation of the Fisher information for the spin-boson model. To perform the simulations we made use of the recently developed tensor-network TEMPO algorithm and its extension to multi-time measurement scenarios \cite{Strathearn2018NC,Jorgensen2019arxiv}. Details of the simulations are provided in \appref{app:tensor}. Making use of this algorithm has the benefit that the temperature derivative of the excited state can itself be expressed as a tensor network and computed to the same level of accuracy as the probability itself.

Results for the Ohmic and the sub-Ohmic cases are shown in Fig.~\ref{fig:FI_sbm}. Generally we find that the short-time approximation provides a good description of the observed scaling behaviour at sufficiently short times. Even more interesting we note that the scaling behaviour predicted within the short-time approximation
is valid even at times well beyond the regime in which the short-time approximation is expected to hold ($\alpha \delta t^{2} \Gamma(1+s) \omega_{c}^{2} \ll 1$). This indicates that the predicted precision scaling is experimentally relevant, even without the requirement of being able to probe the non-equilibrium qubit dynamics at very short-times. Notice that the low-temperature Fisher information tends to initially increase with time as information about the environment state is extracted by the qubit. After some time the low-temperature Fisher information starts to decrease. This can be understood as the qubit reaching a stationary state, such that a one-time measurement performed on the qubit can no longer probe the relaxation dynamics induced by the coupling with the thermal bath (see also \cite{Correa2015PRL,Mehboudi2019}).

Finally, we note that at sufficiently low temperatures the simulated Fisher information differs from the Markovian result,
even for the rather weak coupling and long times considered here. A similar effect was observed in the context of temperature estimation via the Kubo-Martin-Schwinger-like relations obeyed by emission and absorption spectra of multichromophoric systems \cite{Buser_PRA2017_InitialCorrelations}.
There it was pointed out that faithfully recovering the temperature from observed spectra
requires taking into account system-environment correlations. This is true even at very low coupling strengths, where these correlations are generally weak.

%%%%%%%%%%%%%%%%%%%%%%%%%%%%%%%%%%%%%%%%%%%%%%%

\section{Conclusion}
In this paper we have discussed precision scaling for thermometry in cold quantum systems. In particular, we have investigated how finite measurement resolution, meaning that states that are close in energy cannot be perfectly distinguished, impacts the precision scaling. We have proposed a finite-resolution criterion characterising such measurements. Based on this, we derived a tightened bound on the scaling of the Fisher information. Furthermore, we showed that this bound is tight as it can be saturated via both an exponential resolution measurement as well as an excitation-preserving, single-qubit measurement on a sample of non-interacting bosons. We validated the approximations involved in demonstrating tightness for the single-qubit measurement by performing a numerical simulation of the sub-Ohmic spin-boson model. Here, we provided an illustration of a Fisher information scaling linearly with temperature. Interestingly, as far as we are aware, this is the best scaling which has been found in any concrete physical model subject to finite-resolution constraints.

\begin{acknowledgments}
MRJ and JBB were supported by the Independent Research Fund Denmark. PPP acknowledges funding from the European Union's Horizon 2020 research and innovation programme under the Marie Sk{\l}odowska-Curie Grant Agreement No. 796700.
\end{acknowledgments}

\bibliography{bibliography.bib}

%merlin.mbs apsrev4-1.bst 2010-07-25 4.21a (PWD, AO, DPC) hacked
%Control: key (0)
%Control: author (0) dotless jnrlst
%Control: editor formatted (1) identically to author
%Control: production of article title (0) allowed
%Control: page (1) range
%Control: year (0) verbatim
%Control: production of eprint (0) enabled
\begin{thebibliography}{53}%
\makeatletter
\providecommand \@ifxundefined [1]{%
 \@ifx{#1\undefined}
}%
\providecommand \@ifnum [1]{%
 \ifnum #1\expandafter \@firstoftwo
 \else \expandafter \@secondoftwo
 \fi
}%
\providecommand \@ifx [1]{%
 \ifx #1\expandafter \@firstoftwo
 \else \expandafter \@secondoftwo
 \fi
}%
\providecommand \natexlab [1]{#1}%
\providecommand \enquote  [1]{``#1''}%
\providecommand \bibnamefont  [1]{#1}%
\providecommand \bibfnamefont [1]{#1}%
\providecommand \citenamefont [1]{#1}%
\providecommand \href@noop [0]{\@secondoftwo}%
\providecommand \href [0]{\begingroup \@sanitize@url \@href}%
\providecommand \@href[1]{\@@startlink{#1}\@@href}%
\providecommand \@@href[1]{\endgroup#1\@@endlink}%
\providecommand \@sanitize@url [0]{\catcode `\\12\catcode `\$12\catcode
  `\&12\catcode `\#12\catcode `\^12\catcode `\_12\catcode `\%12\relax}%
\providecommand \@@startlink[1]{}%
\providecommand \@@endlink[0]{}%
\providecommand \url  [0]{\begingroup\@sanitize@url \@url }%
\providecommand \@url [1]{\endgroup\@href {#1}{\urlprefix }}%
\providecommand \urlprefix  [0]{URL }%
\providecommand \Eprint [0]{\href }%
\providecommand \doibase [0]{http://dx.doi.org/}%
\providecommand \selectlanguage [0]{\@gobble}%
\providecommand \bibinfo  [0]{\@secondoftwo}%
\providecommand \bibfield  [0]{\@secondoftwo}%
\providecommand \translation [1]{[#1]}%
\providecommand \BibitemOpen [0]{}%
\providecommand \bibitemStop [0]{}%
\providecommand \bibitemNoStop [0]{.\EOS\space}%
\providecommand \EOS [0]{\spacefactor3000\relax}%
\providecommand \BibitemShut  [1]{\csname bibitem#1\endcsname}%
\let\auto@bib@innerbib\@empty
%</preamble>
\bibitem [{\citenamefont {Carlos}\ and\ \citenamefont
  {Palacio}(2016)}]{Carlos2016}%
  \BibitemOpen
  \bibinfo {editor} {\bibfnamefont {Lu\'is~Dias}\ \bibnamefont {Carlos}}\ and\
  \bibinfo {editor} {\bibfnamefont {Fernando}\ \bibnamefont {Palacio}},\ eds.,\
  \href {\doibase 10.1039/9781782622031} {\emph {\bibinfo {title} {Thermometry
  at the Nanoscale}}}\ (\bibinfo  {publisher} {The Royal Society of
  Chemistry},\ \bibinfo {year} {2016})\ pp.\ \bibinfo {pages}
  {P007--522}\BibitemShut {NoStop}%
\bibitem [{\citenamefont {Mehboudi}\ \emph
  {et~al.}(2019{\natexlab{a}})\citenamefont {Mehboudi}, \citenamefont
  {Sanpera},\ and\ \citenamefont {Correa}}]{Mehboudi2019}%
  \BibitemOpen
  \bibfield  {author} {\bibinfo {author} {\bibfnamefont {Mohammad}\
  \bibnamefont {Mehboudi}}, \bibinfo {author} {\bibfnamefont {Anna}\
  \bibnamefont {Sanpera}}, \ and\ \bibinfo {author} {\bibfnamefont {Luis~A}\
  \bibnamefont {Correa}},\ }\bibfield  {title} {\enquote {\bibinfo {title}
  {Thermometry in the quantum regime: recent theoretical progress},}\ }\href
  {\doibase 10.1088/1751-8121/ab2828} {\bibfield  {journal} {\bibinfo
  {journal} {Journal of Physics A: Mathematical and Theoretical}\ }\textbf
  {\bibinfo {volume} {52}},\ \bibinfo {pages} {303001} (\bibinfo {year}
  {2019}{\natexlab{a}})}\BibitemShut {NoStop}%
\bibitem [{\citenamefont {Weld}\ \emph {et~al.}(2009)\citenamefont {Weld},
  \citenamefont {Medley}, \citenamefont {Miyake}, \citenamefont {Hucul},
  \citenamefont {Pritchard},\ and\ \citenamefont
  {Ketterle}}]{PhysRevLett.103.245301}%
  \BibitemOpen
  \bibfield  {author} {\bibinfo {author} {\bibfnamefont {David~M.}\
  \bibnamefont {Weld}}, \bibinfo {author} {\bibfnamefont {Patrick}\
  \bibnamefont {Medley}}, \bibinfo {author} {\bibfnamefont {Hirokazu}\
  \bibnamefont {Miyake}}, \bibinfo {author} {\bibfnamefont {David}\
  \bibnamefont {Hucul}}, \bibinfo {author} {\bibfnamefont {David~E.}\
  \bibnamefont {Pritchard}}, \ and\ \bibinfo {author} {\bibfnamefont
  {Wolfgang}\ \bibnamefont {Ketterle}},\ }\bibfield  {title} {\enquote
  {\bibinfo {title} {Spin gradient thermometry for ultracold atoms in optical
  lattices},}\ }\href {\doibase 10.1103/PhysRevLett.103.245301} {\bibfield
  {journal} {\bibinfo  {journal} {Phys. Rev. Lett.}\ }\textbf {\bibinfo
  {volume} {103}},\ \bibinfo {pages} {245301} (\bibinfo {year}
  {2009})}\BibitemShut {NoStop}%
\bibitem [{\citenamefont {Gasparinetti}\ \emph {et~al.}(2011)\citenamefont
  {Gasparinetti}, \citenamefont {Deon}, \citenamefont {Biasiol}, \citenamefont
  {Sorba}, \citenamefont {Beltram},\ and\ \citenamefont
  {Giazotto}}]{Gasparinetti2011}%
  \BibitemOpen
  \bibfield  {author} {\bibinfo {author} {\bibfnamefont {S.}~\bibnamefont
  {Gasparinetti}}, \bibinfo {author} {\bibfnamefont {F.}~\bibnamefont {Deon}},
  \bibinfo {author} {\bibfnamefont {G.}~\bibnamefont {Biasiol}}, \bibinfo
  {author} {\bibfnamefont {L.}~\bibnamefont {Sorba}}, \bibinfo {author}
  {\bibfnamefont {F.}~\bibnamefont {Beltram}}, \ and\ \bibinfo {author}
  {\bibfnamefont {F.}~\bibnamefont {Giazotto}},\ }\bibfield  {title} {\enquote
  {\bibinfo {title} {Probing the local temperature of a two-dimensional
  electron gas microdomain with a quantum dot: Measurement of electron-phonon
  interaction},}\ }\href {\doibase 10.1103/PhysRevB.83.201306} {\bibfield
  {journal} {\bibinfo  {journal} {Phys. Rev. B}\ }\textbf {\bibinfo {volume}
  {83}},\ \bibinfo {pages} {201306(R)} (\bibinfo {year} {2011})}\BibitemShut
  {NoStop}%
\bibitem [{\citenamefont {Neumann}\ \emph {et~al.}(2013)\citenamefont
  {Neumann}, \citenamefont {Jakobi}, \citenamefont {Dolde}, \citenamefont
  {Burk}, \citenamefont {Reuter}, \citenamefont {Waldherr}, \citenamefont
  {Honert}, \citenamefont {Wolf}, \citenamefont {Brunner}, \citenamefont
  {Shim}, \citenamefont {Suter}, \citenamefont {Sumiya}, \citenamefont
  {Isoya},\ and\ \citenamefont {Wrachtrup}}]{Neumann2013NanoLett}%
  \BibitemOpen
  \bibfield  {author} {\bibinfo {author} {\bibfnamefont {P.}~\bibnamefont
  {Neumann}}, \bibinfo {author} {\bibfnamefont {I.}~\bibnamefont {Jakobi}},
  \bibinfo {author} {\bibfnamefont {F.}~\bibnamefont {Dolde}}, \bibinfo
  {author} {\bibfnamefont {C.}~\bibnamefont {Burk}}, \bibinfo {author}
  {\bibfnamefont {R.}~\bibnamefont {Reuter}}, \bibinfo {author} {\bibfnamefont
  {G.}~\bibnamefont {Waldherr}}, \bibinfo {author} {\bibfnamefont
  {J.}~\bibnamefont {Honert}}, \bibinfo {author} {\bibfnamefont
  {T.}~\bibnamefont {Wolf}}, \bibinfo {author} {\bibfnamefont {A.}~\bibnamefont
  {Brunner}}, \bibinfo {author} {\bibfnamefont {J.~H.}\ \bibnamefont {Shim}},
  \bibinfo {author} {\bibfnamefont {D.}~\bibnamefont {Suter}}, \bibinfo
  {author} {\bibfnamefont {H.}~\bibnamefont {Sumiya}}, \bibinfo {author}
  {\bibfnamefont {J.}~\bibnamefont {Isoya}}, \ and\ \bibinfo {author}
  {\bibfnamefont {J.}~\bibnamefont {Wrachtrup}},\ }\bibfield  {title} {\enquote
  {\bibinfo {title} {High-precision nanoscale temperature sensing using single
  defects in diamond},}\ }\href {\doibase 10.1021/nl401216y} {\bibfield
  {journal} {\bibinfo  {journal} {Nano Letters}\ }\textbf {\bibinfo {volume}
  {13}},\ \bibinfo {pages} {2738--2742} (\bibinfo {year} {2013})}\BibitemShut
  {NoStop}%
\bibitem [{\citenamefont {Kucsko}\ \emph {et~al.}(2013)\citenamefont {Kucsko},
  \citenamefont {Maurer}, \citenamefont {Yao}, \citenamefont {Kubo},
  \citenamefont {Noh}, \citenamefont {Lo}, \citenamefont {Park},\ and\
  \citenamefont {Lukin}}]{Lukin2013Nature}%
  \BibitemOpen
  \bibfield  {author} {\bibinfo {author} {\bibfnamefont {G.}~\bibnamefont
  {Kucsko}}, \bibinfo {author} {\bibfnamefont {P.~C.}\ \bibnamefont {Maurer}},
  \bibinfo {author} {\bibfnamefont {N.~Y.}\ \bibnamefont {Yao}}, \bibinfo
  {author} {\bibfnamefont {M.}~\bibnamefont {Kubo}}, \bibinfo {author}
  {\bibfnamefont {N.~J.}\ \bibnamefont {Noh}}, \bibinfo {author} {\bibfnamefont
  {P.~K.}\ \bibnamefont {Lo}}, \bibinfo {author} {\bibfnamefont
  {H.}~\bibnamefont {Park}}, \ and\ \bibinfo {author} {\bibfnamefont {M.~D.}\
  \bibnamefont {Lukin}},\ }\bibfield  {title} {\enquote {\bibinfo {title}
  {Nanometre-scale thermometry in a living cell},}\ }\href {\doibase
  10.1038/nature12373} {\bibfield  {journal} {\bibinfo  {journal} {Nature
  Letter}\ }\textbf {\bibinfo {volume} {500}},\ \bibinfo {pages} {54--58}
  (\bibinfo {year} {2013})}\BibitemShut {NoStop}%
\bibitem [{\citenamefont {Sabin}\ \emph {et~al.}(2014)\citenamefont {Sabin},
  \citenamefont {White}, \citenamefont {Hackermuller},\ and\ \citenamefont
  {Fuentes}}]{Sabin2014ScientificReports}%
  \BibitemOpen
  \bibfield  {author} {\bibinfo {author} {\bibfnamefont {Carlos}\ \bibnamefont
  {Sabin}}, \bibinfo {author} {\bibfnamefont {Angela}\ \bibnamefont {White}},
  \bibinfo {author} {\bibfnamefont {Lucia}\ \bibnamefont {Hackermuller}}, \
  and\ \bibinfo {author} {\bibfnamefont {Ivette}\ \bibnamefont {Fuentes}},\
  }\bibfield  {title} {\enquote {\bibinfo {title} {Impurities as a quantum
  thermometer for a bose-einstein condensate},}\ }\href {\doibase
  10.1038/srep06436} {\bibfield  {journal} {\bibinfo  {journal} {Scientific
  Reports}\ }\textbf {\bibinfo {volume} {4}},\ \bibinfo {pages} {6436}
  (\bibinfo {year} {2014})}\BibitemShut {NoStop}%
\bibitem [{\citenamefont {Gasparinetti}\ \emph {et~al.}(2015)\citenamefont
  {Gasparinetti}, \citenamefont {Viisanen}, \citenamefont {Saira},
  \citenamefont {Faivre}, \citenamefont {Arzeo}, \citenamefont {Meschke},\ and\
  \citenamefont {Pekola}}]{Gasparinetti2015}%
  \BibitemOpen
  \bibfield  {author} {\bibinfo {author} {\bibfnamefont {S.}~\bibnamefont
  {Gasparinetti}}, \bibinfo {author} {\bibfnamefont {K.~L.}\ \bibnamefont
  {Viisanen}}, \bibinfo {author} {\bibfnamefont {O.-P.}\ \bibnamefont {Saira}},
  \bibinfo {author} {\bibfnamefont {T.}~\bibnamefont {Faivre}}, \bibinfo
  {author} {\bibfnamefont {M.}~\bibnamefont {Arzeo}}, \bibinfo {author}
  {\bibfnamefont {M.}~\bibnamefont {Meschke}}, \ and\ \bibinfo {author}
  {\bibfnamefont {J.~P.}\ \bibnamefont {Pekola}},\ }\bibfield  {title}
  {\enquote {\bibinfo {title} {Fast electron thermometry for ultrasensitive
  calorimetric detection},}\ }\href {\doibase 10.1103/PhysRevApplied.3.014007}
  {\bibfield  {journal} {\bibinfo  {journal} {Phys. Rev. Applied}\ }\textbf
  {\bibinfo {volume} {3}},\ \bibinfo {pages} {014007} (\bibinfo {year}
  {2015})}\BibitemShut {NoStop}%
\bibitem [{\citenamefont {Moldover}\ \emph {et~al.}(2016)\citenamefont
  {Moldover}, \citenamefont {Tew},\ and\ \citenamefont {Yoon}}]{Moldover2016}%
  \BibitemOpen
  \bibfield  {author} {\bibinfo {author} {\bibfnamefont {Michael~R.}\
  \bibnamefont {Moldover}}, \bibinfo {author} {\bibfnamefont {Weston~L.}\
  \bibnamefont {Tew}}, \ and\ \bibinfo {author} {\bibfnamefont {Howard~W.}\
  \bibnamefont {Yoon}},\ }\bibfield  {title} {\enquote {\bibinfo {title}
  {Advances in thermometry},}\ }\href {\doibase 10.1038/nphys3618} {\bibfield
  {journal} {\bibinfo  {journal} {Nature Physics}\ }\textbf {\bibinfo {volume}
  {12}},\ \bibinfo {pages} {7} (\bibinfo {year} {2016})}\BibitemShut {NoStop}%
\bibitem [{\citenamefont {Palma}\ \emph {et~al.}(2017)\citenamefont {Palma},
  \citenamefont {Scheller}, \citenamefont {Maradan}, \citenamefont
  {Feshchenko}, \citenamefont {Meschke},\ and\ \citenamefont
  {Zumbuhl}}]{Palma2017ChipCooling}%
  \BibitemOpen
  \bibfield  {author} {\bibinfo {author} {\bibfnamefont {M.}~\bibnamefont
  {Palma}}, \bibinfo {author} {\bibfnamefont {C.~P.}\ \bibnamefont {Scheller}},
  \bibinfo {author} {\bibfnamefont {D.}~\bibnamefont {Maradan}}, \bibinfo
  {author} {\bibfnamefont {A.~V.}\ \bibnamefont {Feshchenko}}, \bibinfo
  {author} {\bibfnamefont {M.}~\bibnamefont {Meschke}}, \ and\ \bibinfo
  {author} {\bibfnamefont {D.~M.}\ \bibnamefont {Zumbuhl}},\ }\bibfield
  {title} {\enquote {\bibinfo {title} {On-and-off chip cooling of a coulomb
  blockade thermometer down to 2.8 mk},}\ }\href {\doibase 10.1063/1.5002565}
  {\bibfield  {journal} {\bibinfo  {journal} {Applied Physics Letters}\
  }\textbf {\bibinfo {volume} {111}},\ \bibinfo {pages} {253105} (\bibinfo
  {year} {2017})}\BibitemShut {NoStop}%
\bibitem [{\citenamefont {Zgirski}\ \emph {et~al.}(2018)\citenamefont
  {Zgirski}, \citenamefont {Foltyn}, \citenamefont {Savin}, \citenamefont
  {Norowski}, \citenamefont {Meschke},\ and\ \citenamefont
  {Pekola}}]{PhysRevApplied.10.044068}%
  \BibitemOpen
  \bibfield  {author} {\bibinfo {author} {\bibfnamefont {M.}~\bibnamefont
  {Zgirski}}, \bibinfo {author} {\bibfnamefont {M.}~\bibnamefont {Foltyn}},
  \bibinfo {author} {\bibfnamefont {A.}~\bibnamefont {Savin}}, \bibinfo
  {author} {\bibfnamefont {K.}~\bibnamefont {Norowski}}, \bibinfo {author}
  {\bibfnamefont {M.}~\bibnamefont {Meschke}}, \ and\ \bibinfo {author}
  {\bibfnamefont {J.}~\bibnamefont {Pekola}},\ }\bibfield  {title} {\enquote
  {\bibinfo {title} {Nanosecond thermometry with josephson junctions},}\ }\href
  {\doibase 10.1103/PhysRevApplied.10.044068} {\bibfield  {journal} {\bibinfo
  {journal} {Phys. Rev. Applied}\ }\textbf {\bibinfo {volume} {10}},\ \bibinfo
  {pages} {044068} (\bibinfo {year} {2018})}\BibitemShut {NoStop}%
\bibitem [{\citenamefont {Scigliuzzo}\ \emph {et~al.}(2020)\citenamefont
  {Scigliuzzo}, \citenamefont {Bengtsson}, \citenamefont {Besse}, \citenamefont
  {Wallraff}, \citenamefont {Delsing},\ and\ \citenamefont
  {Gasparinetti}}]{Scigliuzzo2020}%
  \BibitemOpen
  \bibfield  {author} {\bibinfo {author} {\bibfnamefont {Marco}\ \bibnamefont
  {Scigliuzzo}}, \bibinfo {author} {\bibfnamefont {Andreas}\ \bibnamefont
  {Bengtsson}}, \bibinfo {author} {\bibfnamefont {Jean-Claude}\ \bibnamefont
  {Besse}}, \bibinfo {author} {\bibfnamefont {Andreas}\ \bibnamefont
  {Wallraff}}, \bibinfo {author} {\bibfnamefont {Per}\ \bibnamefont {Delsing}},
  \ and\ \bibinfo {author} {\bibfnamefont {Simone}\ \bibnamefont
  {Gasparinetti}},\ }\bibfield  {title} {\enquote {\bibinfo {title} {Primary
  thermometry of propagating microwaves in the quantum regime},}\ }\href
  {https://arxiv.org/abs/2003.13522} {\bibfield  {journal} {\bibinfo  {journal}
  {arXiv:2003:13522v1 [quant-ph]}\ } (\bibinfo {year} {2020})}\BibitemShut
  {NoStop}%
\bibitem [{\citenamefont {Correa}\ \emph {et~al.}(2015)\citenamefont {Correa},
  \citenamefont {Mehboudi}, \citenamefont {Adesso},\ and\ \citenamefont
  {Sanpera}}]{Correa2015PRL}%
  \BibitemOpen
  \bibfield  {author} {\bibinfo {author} {\bibfnamefont {Luis~A.}\ \bibnamefont
  {Correa}}, \bibinfo {author} {\bibfnamefont {Mohammad}\ \bibnamefont
  {Mehboudi}}, \bibinfo {author} {\bibfnamefont {Gerardo}\ \bibnamefont
  {Adesso}}, \ and\ \bibinfo {author} {\bibfnamefont {Anna}\ \bibnamefont
  {Sanpera}},\ }\bibfield  {title} {\enquote {\bibinfo {title} {Individual
  quantum probes for optimal thermometry},}\ }\href {\doibase
  10.1103/PhysRevLett.114.220405} {\bibfield  {journal} {\bibinfo  {journal}
  {Phys. Rev. Lett.}\ }\textbf {\bibinfo {volume} {114}},\ \bibinfo {pages}
  {220405} (\bibinfo {year} {2015})}\BibitemShut {NoStop}%
\bibitem [{\citenamefont {Correa}\ \emph {et~al.}(2017)\citenamefont {Correa},
  \citenamefont {Perarnau-Llobet}, \citenamefont {Hovhannisyan}, \citenamefont
  {Hern\'andez-Santana}, \citenamefont {Mehboudi},\ and\ \citenamefont
  {Sanpera}}]{PhysRevA.96.062103}%
  \BibitemOpen
  \bibfield  {author} {\bibinfo {author} {\bibfnamefont {Luis~A.}\ \bibnamefont
  {Correa}}, \bibinfo {author} {\bibfnamefont {Mart\'{\i}}\ \bibnamefont
  {Perarnau-Llobet}}, \bibinfo {author} {\bibfnamefont {Karen~V.}\ \bibnamefont
  {Hovhannisyan}}, \bibinfo {author} {\bibfnamefont {Senaida}\ \bibnamefont
  {Hern\'andez-Santana}}, \bibinfo {author} {\bibfnamefont {Mohammad}\
  \bibnamefont {Mehboudi}}, \ and\ \bibinfo {author} {\bibfnamefont {Anna}\
  \bibnamefont {Sanpera}},\ }\bibfield  {title} {\enquote {\bibinfo {title}
  {Enhancement of low-temperature thermometry by strong coupling},}\ }\href
  {\doibase 10.1103/PhysRevA.96.062103} {\bibfield  {journal} {\bibinfo
  {journal} {Phys. Rev. A}\ }\textbf {\bibinfo {volume} {96}},\ \bibinfo
  {pages} {062103} (\bibinfo {year} {2017})}\BibitemShut {NoStop}%
\bibitem [{\citenamefont {Mehboudi}\ \emph
  {et~al.}(2019{\natexlab{b}})\citenamefont {Mehboudi}, \citenamefont {Lampo},
  \citenamefont {Charalambous}, \citenamefont {Correa}, \citenamefont
  {Garcia-March},\ and\ \citenamefont {Lewenstein}}]{Mehboudi2019PRL}%
  \BibitemOpen
  \bibfield  {author} {\bibinfo {author} {\bibfnamefont {Mohammad}\
  \bibnamefont {Mehboudi}}, \bibinfo {author} {\bibfnamefont {Aniello}\
  \bibnamefont {Lampo}}, \bibinfo {author} {\bibfnamefont {Christos}\
  \bibnamefont {Charalambous}}, \bibinfo {author} {\bibfnamefont {Luis~A.}\
  \bibnamefont {Correa}}, \bibinfo {author} {\bibfnamefont {Miguel~A.}\
  \bibnamefont {Garcia-March}}, \ and\ \bibinfo {author} {\bibfnamefont
  {Marciej}\ \bibnamefont {Lewenstein}},\ }\bibfield  {title} {\enquote
  {\bibinfo {title} {Using polarons for sub-nk quantum nondemolition
  thermometry in a bose-einstein condensate},}\ }\href {\doibase
  10.1103/PhysRevLett.122.030403} {\bibfield  {journal} {\bibinfo  {journal}
  {Phys. Rev. Lett.}\ }\textbf {\bibinfo {volume} {122}},\ \bibinfo {pages}
  {030403} (\bibinfo {year} {2019}{\natexlab{b}})}\BibitemShut {NoStop}%
\bibitem [{\citenamefont {Hofer}\ \emph {et~al.}(2017)\citenamefont {Hofer},
  \citenamefont {Brask}, \citenamefont {Perarnau-Llobet},\ and\ \citenamefont
  {Brunner}}]{PhysRevLett.119.090603}%
  \BibitemOpen
  \bibfield  {author} {\bibinfo {author} {\bibfnamefont {Patrick~P.}\
  \bibnamefont {Hofer}}, \bibinfo {author} {\bibfnamefont {Jonatan~Bohr}\
  \bibnamefont {Brask}}, \bibinfo {author} {\bibfnamefont {Mart\'{\i}}\
  \bibnamefont {Perarnau-Llobet}}, \ and\ \bibinfo {author} {\bibfnamefont
  {Nicolas}\ \bibnamefont {Brunner}},\ }\bibfield  {title} {\enquote {\bibinfo
  {title} {Quantum thermal machine as a thermometer},}\ }\href {\doibase
  10.1103/PhysRevLett.119.090603} {\bibfield  {journal} {\bibinfo  {journal}
  {Phys. Rev. Lett.}\ }\textbf {\bibinfo {volume} {119}},\ \bibinfo {pages}
  {090603} (\bibinfo {year} {2017})}\BibitemShut {NoStop}%
\bibitem [{\citenamefont {Guarnieri}\ \emph {et~al.}(2019)\citenamefont
  {Guarnieri}, \citenamefont {Landi}, \citenamefont {Clark},\ and\
  \citenamefont {Goold}}]{Guarnieri2019}%
  \BibitemOpen
  \bibfield  {author} {\bibinfo {author} {\bibfnamefont {Giacomo}\ \bibnamefont
  {Guarnieri}}, \bibinfo {author} {\bibfnamefont {Gabriel~T.}\ \bibnamefont
  {Landi}}, \bibinfo {author} {\bibfnamefont {Stephen~R.}\ \bibnamefont
  {Clark}}, \ and\ \bibinfo {author} {\bibfnamefont {John}\ \bibnamefont
  {Goold}},\ }\bibfield  {title} {\enquote {\bibinfo {title} {Thermodynamics of
  precision in quantum nonequilibrium steady states},}\ }\href {\doibase
  10.1103/PhysRevResearch.1.033021} {\bibfield  {journal} {\bibinfo  {journal}
  {Phys. Rev. Research}\ }\textbf {\bibinfo {volume} {1}},\ \bibinfo {pages}
  {033021} (\bibinfo {year} {2019})}\BibitemShut {NoStop}%
\bibitem [{\citenamefont {Latune}\ \emph {et~al.}(2020)\citenamefont {Latune},
  \citenamefont {Sinayskiy},\ and\ \citenamefont {Petruccione}}]{Latune2020}%
  \BibitemOpen
  \bibfield  {author} {\bibinfo {author} {\bibfnamefont {Camille~L.}\
  \bibnamefont {Latune}}, \bibinfo {author} {\bibfnamefont {Ilya}\ \bibnamefont
  {Sinayskiy}}, \ and\ \bibinfo {author} {\bibfnamefont {Francesco}\
  \bibnamefont {Petruccione}},\ }\bibfield  {title} {\enquote {\bibinfo {title}
  {Collective heat capacity for quantum thermometry and quantum engine
  enhancements},}\ }\href {https://arxiv.org/abs/2004.00032} {\bibfield
  {journal} {\bibinfo  {journal} {arXiv:2004.00032 [quant-ph]}\ } (\bibinfo
  {year} {2020})}\BibitemShut {NoStop}%
\bibitem [{\citenamefont {Jevtic}\ \emph {et~al.}(2015)\citenamefont {Jevtic},
  \citenamefont {Newman}, \citenamefont {Rudolph},\ and\ \citenamefont
  {Stace}}]{PhysRevA.91.012331}%
  \BibitemOpen
  \bibfield  {author} {\bibinfo {author} {\bibfnamefont {Sania}\ \bibnamefont
  {Jevtic}}, \bibinfo {author} {\bibfnamefont {David}\ \bibnamefont {Newman}},
  \bibinfo {author} {\bibfnamefont {Terry}\ \bibnamefont {Rudolph}}, \ and\
  \bibinfo {author} {\bibfnamefont {T.~M.}\ \bibnamefont {Stace}},\ }\bibfield
  {title} {\enquote {\bibinfo {title} {Single-qubit thermometry},}\ }\href
  {\doibase 10.1103/PhysRevA.91.012331} {\bibfield  {journal} {\bibinfo
  {journal} {Phys. Rev. A}\ }\textbf {\bibinfo {volume} {91}},\ \bibinfo
  {pages} {012331} (\bibinfo {year} {2015})}\BibitemShut {NoStop}%
\bibitem [{\citenamefont {Mancino}\ \emph {et~al.}(2017)\citenamefont
  {Mancino}, \citenamefont {Sbroscia}, \citenamefont {Gianani}, \citenamefont
  {Roccia},\ and\ \citenamefont {Barbieri}}]{PhysRevLett.118.130502}%
  \BibitemOpen
  \bibfield  {author} {\bibinfo {author} {\bibfnamefont {Luca}\ \bibnamefont
  {Mancino}}, \bibinfo {author} {\bibfnamefont {Marco}\ \bibnamefont
  {Sbroscia}}, \bibinfo {author} {\bibfnamefont {Ilaria}\ \bibnamefont
  {Gianani}}, \bibinfo {author} {\bibfnamefont {Emanuele}\ \bibnamefont
  {Roccia}}, \ and\ \bibinfo {author} {\bibfnamefont {Marco}\ \bibnamefont
  {Barbieri}},\ }\bibfield  {title} {\enquote {\bibinfo {title} {Quantum
  simulation of single-qubit thermometry using linear optics},}\ }\href
  {\doibase 10.1103/PhysRevLett.118.130502} {\bibfield  {journal} {\bibinfo
  {journal} {Phys. Rev. Lett.}\ }\textbf {\bibinfo {volume} {118}},\ \bibinfo
  {pages} {130502} (\bibinfo {year} {2017})}\BibitemShut {NoStop}%
\bibitem [{\citenamefont {Johnson}\ \emph {et~al.}(2016)\citenamefont
  {Johnson}, \citenamefont {Cosco}, \citenamefont {Mitchison}, \citenamefont
  {Jaksch},\ and\ \citenamefont {Clark}}]{PhysRevA.93.053619}%
  \BibitemOpen
  \bibfield  {author} {\bibinfo {author} {\bibfnamefont {T.~H.}\ \bibnamefont
  {Johnson}}, \bibinfo {author} {\bibfnamefont {F.}~\bibnamefont {Cosco}},
  \bibinfo {author} {\bibfnamefont {M.~T.}\ \bibnamefont {Mitchison}}, \bibinfo
  {author} {\bibfnamefont {D.}~\bibnamefont {Jaksch}}, \ and\ \bibinfo {author}
  {\bibfnamefont {S.~R.}\ \bibnamefont {Clark}},\ }\bibfield  {title} {\enquote
  {\bibinfo {title} {Thermometry of ultracold atoms via nonequilibrium work
  distributions},}\ }\href {\doibase 10.1103/PhysRevA.93.053619} {\bibfield
  {journal} {\bibinfo  {journal} {Phys. Rev. A}\ }\textbf {\bibinfo {volume}
  {93}},\ \bibinfo {pages} {053619} (\bibinfo {year} {2016})}\BibitemShut
  {NoStop}%
\bibitem [{\citenamefont {Kiilerich}\ \emph {et~al.}(2018)\citenamefont
  {Kiilerich}, \citenamefont {De~Pasquale},\ and\ \citenamefont
  {Giovannetti}}]{Killerich2018PRA}%
  \BibitemOpen
  \bibfield  {author} {\bibinfo {author} {\bibfnamefont {Alexander~H.}\
  \bibnamefont {Kiilerich}}, \bibinfo {author} {\bibfnamefont {Antonella}\
  \bibnamefont {De~Pasquale}}, \ and\ \bibinfo {author} {\bibfnamefont
  {Vittorio}\ \bibnamefont {Giovannetti}},\ }\bibfield  {title} {\enquote
  {\bibinfo {title} {Dynamical approach to ancilla-assisted quantum
  thermometry},}\ }\href {\doibase 10.1103/PhysRevA.98.042124} {\bibfield
  {journal} {\bibinfo  {journal} {Phys. Rev. A}\ }\textbf {\bibinfo {volume}
  {98}},\ \bibinfo {pages} {042124} (\bibinfo {year} {2018})}\BibitemShut
  {NoStop}%
\bibitem [{\citenamefont {Feyles}\ \emph {et~al.}(2019)\citenamefont {Feyles},
  \citenamefont {Mancino}, \citenamefont {Sbroscia}, \citenamefont {Gianani},\
  and\ \citenamefont {Barbieri}}]{Feyles2019PRA}%
  \BibitemOpen
  \bibfield  {author} {\bibinfo {author} {\bibfnamefont {Michele~M.}\
  \bibnamefont {Feyles}}, \bibinfo {author} {\bibfnamefont {Luca}\ \bibnamefont
  {Mancino}}, \bibinfo {author} {\bibfnamefont {Marco}\ \bibnamefont
  {Sbroscia}}, \bibinfo {author} {\bibfnamefont {Ilaria}\ \bibnamefont
  {Gianani}}, \ and\ \bibinfo {author} {\bibfnamefont {Marco}\ \bibnamefont
  {Barbieri}},\ }\bibfield  {title} {\enquote {\bibinfo {title} {Dynamical role
  of quantum signatures in quantum thermometry},}\ }\href {\doibase
  10.1103/PhysRevA.99.062114} {\bibfield  {journal} {\bibinfo  {journal} {Phys.
  Rev. A}\ }\textbf {\bibinfo {volume} {99}},\ \bibinfo {pages} {062114}
  (\bibinfo {year} {2019})}\BibitemShut {NoStop}%
\bibitem [{\citenamefont {Mukherjee}\ \emph {et~al.}(2019)\citenamefont
  {Mukherjee}, \citenamefont {Zwick}, \citenamefont {Ghosh}, \citenamefont
  {Chen},\ and\ \citenamefont {Kurizki}}]{Mukherjee2019}%
  \BibitemOpen
  \bibfield  {author} {\bibinfo {author} {\bibfnamefont {Victor}\ \bibnamefont
  {Mukherjee}}, \bibinfo {author} {\bibfnamefont {Analia}\ \bibnamefont
  {Zwick}}, \bibinfo {author} {\bibfnamefont {Arnab}\ \bibnamefont {Ghosh}},
  \bibinfo {author} {\bibfnamefont {Xi}~\bibnamefont {Chen}}, \ and\ \bibinfo
  {author} {\bibfnamefont {Gershon}\ \bibnamefont {Kurizki}},\ }\bibfield
  {title} {\enquote {\bibinfo {title} {Enhanced precision bound of
  low-temperature quantum thermometry via dynamical control},}\ }\href
  {https://doi.org/10.1038/s42005-019-0265-y} {\bibfield  {journal} {\bibinfo
  {journal} {Communications Physics}\ }\textbf {\bibinfo {volume} {2}},\
  \bibinfo {pages} {162} (\bibinfo {year} {2019})}\BibitemShut {NoStop}%
\bibitem [{\citenamefont {Jarzyna}\ and\ \citenamefont
  {Zwierz}(2015)}]{PhysRevA.92.032112}%
  \BibitemOpen
  \bibfield  {author} {\bibinfo {author} {\bibfnamefont {Marcin}\ \bibnamefont
  {Jarzyna}}\ and\ \bibinfo {author} {\bibfnamefont {Marcin}\ \bibnamefont
  {Zwierz}},\ }\bibfield  {title} {\enquote {\bibinfo {title} {Quantum
  interferometric measurements of temperature},}\ }\href {\doibase
  10.1103/PhysRevA.92.032112} {\bibfield  {journal} {\bibinfo  {journal} {Phys.
  Rev. A}\ }\textbf {\bibinfo {volume} {92}},\ \bibinfo {pages} {032112}
  (\bibinfo {year} {2015})}\BibitemShut {NoStop}%
\bibitem [{\citenamefont {Stace}(2010)}]{PhysRevA.82.011611}%
  \BibitemOpen
  \bibfield  {author} {\bibinfo {author} {\bibfnamefont {Thomas~M.}\
  \bibnamefont {Stace}},\ }\bibfield  {title} {\enquote {\bibinfo {title}
  {Quantum limits of thermometry},}\ }\href {\doibase
  10.1103/PhysRevA.82.011611} {\bibfield  {journal} {\bibinfo  {journal} {Phys.
  Rev. A}\ }\textbf {\bibinfo {volume} {82}},\ \bibinfo {pages} {011611(R)}
  (\bibinfo {year} {2010})}\BibitemShut {NoStop}%
\bibitem [{\citenamefont {Cavina}\ \emph {et~al.}(2018)\citenamefont {Cavina},
  \citenamefont {Mancino}, \citenamefont {De~Pasquale}, \citenamefont
  {Gianani}, \citenamefont {Sbroscia}, \citenamefont {Booth}, \citenamefont
  {Roccia}, \citenamefont {Raimondi}, \citenamefont {Giovannetti},\ and\
  \citenamefont {Barbieri}}]{Cavina2018PRA}%
  \BibitemOpen
  \bibfield  {author} {\bibinfo {author} {\bibfnamefont {Vasco}\ \bibnamefont
  {Cavina}}, \bibinfo {author} {\bibfnamefont {Luca}\ \bibnamefont {Mancino}},
  \bibinfo {author} {\bibfnamefont {Antonella}\ \bibnamefont {De~Pasquale}},
  \bibinfo {author} {\bibfnamefont {Ilaria}\ \bibnamefont {Gianani}}, \bibinfo
  {author} {\bibfnamefont {Marco}\ \bibnamefont {Sbroscia}}, \bibinfo {author}
  {\bibfnamefont {Robert~I.}\ \bibnamefont {Booth}}, \bibinfo {author}
  {\bibfnamefont {Emanuele}\ \bibnamefont {Roccia}}, \bibinfo {author}
  {\bibfnamefont {Roberto}\ \bibnamefont {Raimondi}}, \bibinfo {author}
  {\bibfnamefont {Vittorio}\ \bibnamefont {Giovannetti}}, \ and\ \bibinfo
  {author} {\bibfnamefont {Marco}\ \bibnamefont {Barbieri}},\ }\bibfield
  {title} {\enquote {\bibinfo {title} {Bridging thermodynamics and metrology in
  nonequilibrium quantum thermometry},}\ }\href {\doibase
  10.1103/PhysRevA.98.050101} {\bibfield  {journal} {\bibinfo  {journal} {Phys.
  Rev. A}\ }\textbf {\bibinfo {volume} {98}},\ \bibinfo {pages} {050101(R)}
  (\bibinfo {year} {2018})}\BibitemShut {NoStop}%
\bibitem [{\citenamefont {De~Pasquale}\ \emph {et~al.}(2017)\citenamefont
  {De~Pasquale}, \citenamefont {Yuasa},\ and\ \citenamefont
  {Giovannetti}}]{PhysRevA.96.012316}%
  \BibitemOpen
  \bibfield  {author} {\bibinfo {author} {\bibfnamefont {Antonella}\
  \bibnamefont {De~Pasquale}}, \bibinfo {author} {\bibfnamefont {Kazuya}\
  \bibnamefont {Yuasa}}, \ and\ \bibinfo {author} {\bibfnamefont {Vittorio}\
  \bibnamefont {Giovannetti}},\ }\bibfield  {title} {\enquote {\bibinfo {title}
  {Estimating temperature via sequential measurements},}\ }\href {\doibase
  10.1103/PhysRevA.96.012316} {\bibfield  {journal} {\bibinfo  {journal} {Phys.
  Rev. A}\ }\textbf {\bibinfo {volume} {96}},\ \bibinfo {pages} {012316}
  (\bibinfo {year} {2017})}\BibitemShut {NoStop}%
\bibitem [{\citenamefont {Seah}\ \emph {et~al.}(2019)\citenamefont {Seah},
  \citenamefont {Nimmrichter}, \citenamefont {Grimmer}, \citenamefont {Santos},
  \citenamefont {Shu}, \citenamefont {Scarani},\ and\ \citenamefont
  {Landi}}]{Seah2019arxiv}%
  \BibitemOpen
  \bibfield  {author} {\bibinfo {author} {\bibfnamefont {Stella}\ \bibnamefont
  {Seah}}, \bibinfo {author} {\bibfnamefont {Stefan}\ \bibnamefont
  {Nimmrichter}}, \bibinfo {author} {\bibfnamefont {Daniel}\ \bibnamefont
  {Grimmer}}, \bibinfo {author} {\bibfnamefont {Jader~P.}\ \bibnamefont
  {Santos}}, \bibinfo {author} {\bibfnamefont {Aageline}\ \bibnamefont {Shu}},
  \bibinfo {author} {\bibfnamefont {Valerio}\ \bibnamefont {Scarani}}, \ and\
  \bibinfo {author} {\bibfnamefont {Gabriel~T.}\ \bibnamefont {Landi}},\
  }\bibfield  {title} {\enquote {\bibinfo {title} {Collisional quantum
  thermometry},}\ }\href {https://arxiv.org/abs/1904.12551} {\bibfield
  {journal} {\bibinfo  {journal} {arXiv:1904.12551v1 [quant-ph]}\ } (\bibinfo
  {year} {2019})}\BibitemShut {NoStop}%
\bibitem [{\citenamefont {Fr\"owis}\ \emph {et~al.}(2016)\citenamefont
  {Fr\"owis}, \citenamefont {Sekatski},\ and\ \citenamefont
  {D\"ur}}]{Frowis2016PRL}%
  \BibitemOpen
  \bibfield  {author} {\bibinfo {author} {\bibfnamefont {Florian}\ \bibnamefont
  {Fr\"owis}}, \bibinfo {author} {\bibfnamefont {Pavel}\ \bibnamefont
  {Sekatski}}, \ and\ \bibinfo {author} {\bibfnamefont {Wolfgang}\ \bibnamefont
  {D\"ur}},\ }\bibfield  {title} {\enquote {\bibinfo {title} {Detecting large
  quantum fisher information with finite measurement precision},}\ }\href
  {\doibase 10.1103/PhysRevLett.116.090801} {\bibfield  {journal} {\bibinfo
  {journal} {Phys. Rev. Lett.}\ }\textbf {\bibinfo {volume} {116}},\ \bibinfo
  {pages} {090801} (\bibinfo {year} {2016})}\BibitemShut {NoStop}%
\bibitem [{\citenamefont {Potts}\ \emph {et~al.}(2019)\citenamefont {Potts},
  \citenamefont {Brask},\ and\ \citenamefont
  {Brunner}}]{Potts2019fundamentallimits}%
  \BibitemOpen
  \bibfield  {author} {\bibinfo {author} {\bibfnamefont {Patrick~P.}\
  \bibnamefont {Potts}}, \bibinfo {author} {\bibfnamefont {Jonatan~Bohr}\
  \bibnamefont {Brask}}, \ and\ \bibinfo {author} {\bibfnamefont {Nicolas}\
  \bibnamefont {Brunner}},\ }\bibfield  {title} {\enquote {\bibinfo {title}
  {Fundamental limits on low-temperature quantum thermometry with finite
  resolution},}\ }\href {\doibase 10.22331/q-2019-07-09-161} {\bibfield
  {journal} {\bibinfo  {journal} {{Quantum}}\ }\textbf {\bibinfo {volume}
  {3}},\ \bibinfo {pages} {161} (\bibinfo {year} {2019})}\BibitemShut {NoStop}%
\bibitem [{\citenamefont {Guryanova}\ \emph {et~al.}(2020)\citenamefont
  {Guryanova}, \citenamefont {Friis},\ and\ \citenamefont
  {Huber}}]{Guryanova2019arxiv}%
  \BibitemOpen
  \bibfield  {author} {\bibinfo {author} {\bibfnamefont {Yelena}\ \bibnamefont
  {Guryanova}}, \bibinfo {author} {\bibfnamefont {Nicolai}\ \bibnamefont
  {Friis}}, \ and\ \bibinfo {author} {\bibfnamefont {Marcus}\ \bibnamefont
  {Huber}},\ }\bibfield  {title} {\enquote {\bibinfo {title} {Ideal
  {P}rojective {M}easurements {H}ave {I}nfinite {R}esource {C}osts},}\ }\href
  {\doibase 10.22331/q-2020-01-13-222} {\bibfield  {journal} {\bibinfo
  {journal} {{Quantum}}\ }\textbf {\bibinfo {volume} {4}},\ \bibinfo {pages}
  {222} (\bibinfo {year} {2020})}\BibitemShut {NoStop}%
\bibitem [{\citenamefont {De~Pasquale}\ \emph {et~al.}(2015)\citenamefont
  {De~Pasquale}, \citenamefont {Rossini}, \citenamefont {Fazio},\ and\
  \citenamefont {Giovannetti}}]{Pasquale2015Nature}%
  \BibitemOpen
  \bibfield  {author} {\bibinfo {author} {\bibfnamefont {Antonella}\
  \bibnamefont {De~Pasquale}}, \bibinfo {author} {\bibfnamefont {Davide}\
  \bibnamefont {Rossini}}, \bibinfo {author} {\bibfnamefont {Rosario}\
  \bibnamefont {Fazio}}, \ and\ \bibinfo {author} {\bibfnamefont {Vittorio}\
  \bibnamefont {Giovannetti}},\ }\bibfield  {title} {\enquote {\bibinfo {title}
  {Local quantum thermal susceptibility},}\ }\href {\doibase
  10.1038/ncomms12782} {\bibfield  {journal} {\bibinfo  {journal} {Nature
  communications}\ }\textbf {\bibinfo {volume} {7}},\ \bibinfo {pages} {1}
  (\bibinfo {year} {2015})}\BibitemShut {NoStop}%
\bibitem [{\citenamefont {Hovhannisyan}\ and\ \citenamefont
  {Correa}(2018)}]{Hovhannisyan2018PRB}%
  \BibitemOpen
  \bibfield  {author} {\bibinfo {author} {\bibfnamefont {Karen~V.}\
  \bibnamefont {Hovhannisyan}}\ and\ \bibinfo {author} {\bibfnamefont
  {Luis~A.}\ \bibnamefont {Correa}},\ }\bibfield  {title} {\enquote {\bibinfo
  {title} {Measuring the temperature of cold many-body quantum systems},}\
  }\href {\doibase 10.1103/PhysRevB.98.045101} {\bibfield  {journal} {\bibinfo
  {journal} {Phys. Rev. B}\ }\textbf {\bibinfo {volume} {98}},\ \bibinfo
  {pages} {045101} (\bibinfo {year} {2018})}\BibitemShut {NoStop}%
\bibitem [{\citenamefont {De~Palma}\ \emph {et~al.}(2017)\citenamefont
  {De~Palma}, \citenamefont {De~Pasquale},\ and\ \citenamefont
  {Giovannetti}}]{PhysRevA.95.052115}%
  \BibitemOpen
  \bibfield  {author} {\bibinfo {author} {\bibfnamefont {Giacomo}\ \bibnamefont
  {De~Palma}}, \bibinfo {author} {\bibfnamefont {Antonella}\ \bibnamefont
  {De~Pasquale}}, \ and\ \bibinfo {author} {\bibfnamefont {Vittorio}\
  \bibnamefont {Giovannetti}},\ }\bibfield  {title} {\enquote {\bibinfo {title}
  {Universal locality of quantum thermal susceptibility},}\ }\href {\doibase
  10.1103/PhysRevA.95.052115} {\bibfield  {journal} {\bibinfo  {journal} {Phys.
  Rev. A}\ }\textbf {\bibinfo {volume} {95}},\ \bibinfo {pages} {052115}
  (\bibinfo {year} {2017})}\BibitemShut {NoStop}%
\bibitem [{\citenamefont {Razavian}\ \emph {et~al.}(2019)\citenamefont
  {Razavian}, \citenamefont {Benedetti}, \citenamefont {Bina}, \citenamefont
  {Akbari-Kourbolagh},\ and\ \citenamefont {Paris}}]{Razavian2019EPJ}%
  \BibitemOpen
  \bibfield  {author} {\bibinfo {author} {\bibfnamefont {Sholeh}\ \bibnamefont
  {Razavian}}, \bibinfo {author} {\bibfnamefont {Claudia}\ \bibnamefont
  {Benedetti}}, \bibinfo {author} {\bibfnamefont {Matteo}\ \bibnamefont
  {Bina}}, \bibinfo {author} {\bibfnamefont {Yahya}\ \bibnamefont
  {Akbari-Kourbolagh}}, \ and\ \bibinfo {author} {\bibfnamefont {Matteo G.~A.}\
  \bibnamefont {Paris}},\ }\bibfield  {title} {\enquote {\bibinfo {title}
  {Quantum thermometry by single-qubit dephasing},}\ }\href {\doibase
  10.1140/epjp/i2019-12708-9} {\bibfield  {journal} {\bibinfo  {journal} {The
  European Physical Journal Plus}\ }\textbf {\bibinfo {volume} {134}},\
  \bibinfo {pages} {284} (\bibinfo {year} {2019})}\BibitemShut {NoStop}%
\bibitem [{\citenamefont {Paris}(2016)}]{Paris2016JPAMT}%
  \BibitemOpen
  \bibfield  {author} {\bibinfo {author} {\bibfnamefont {Matteo G.~A.}\
  \bibnamefont {Paris}},\ }\bibfield  {title} {\enquote {\bibinfo {title}
  {Achieving the landau bound to precision of quantum thermometry in systems
  with vanishing gap},}\ }\href {\doibase 10.1088/1751-8113/49/3/03LT02}
  {\bibfield  {journal} {\bibinfo  {journal} {J. Phys. A: Math. Theor.}\
  }\textbf {\bibinfo {volume} {49}},\ \bibinfo {pages} {03LT02} (\bibinfo
  {year} {2016})}\BibitemShut {NoStop}%
\bibitem [{\citenamefont {Paris}(2009)}]{Paris2009IJQI}%
  \BibitemOpen
  \bibfield  {author} {\bibinfo {author} {\bibfnamefont {Matteo G.~A.}\
  \bibnamefont {Paris}},\ }\bibfield  {title} {\enquote {\bibinfo {title}
  {Quantum estimation for quantum technology},}\ }\href {\doibase
  10.1142/S0219749909004839} {\bibfield  {journal} {\bibinfo  {journal} {Int.
  J. Quant. Inf.}\ }\textbf {\bibinfo {volume} {7}},\ \bibinfo {pages} {125}
  (\bibinfo {year} {2009})}\BibitemShut {NoStop}%
\bibitem [{\citenamefont {Cramer}(1999)}]{Cramer1999}%
  \BibitemOpen
  \bibfield  {author} {\bibinfo {author} {\bibfnamefont {Harald}\ \bibnamefont
  {Cramer}},\ }\href@noop {} {\emph {\bibinfo {title} {Mathematical methods of
  statistics}}}\ (\bibinfo  {publisher} {Princeton University Press},\ \bibinfo
  {year} {1999})\BibitemShut {NoStop}%
\bibitem [{\citenamefont {Braunstein}\ and\ \citenamefont
  {Caves}(1994)}]{Braunstein1994PRL}%
  \BibitemOpen
  \bibfield  {author} {\bibinfo {author} {\bibfnamefont {S.~L.}\ \bibnamefont
  {Braunstein}}\ and\ \bibinfo {author} {\bibfnamefont {C.~M.}\ \bibnamefont
  {Caves}},\ }\bibfield  {title} {\enquote {\bibinfo {title} {Statistical
  distance and the geometry of quantum states},}\ }\href {\doibase
  10.1103/PhysRevLett.72.3439} {\bibfield  {journal} {\bibinfo  {journal}
  {Phys. Rev. Lett.}\ }\textbf {\bibinfo {volume} {72}},\ \bibinfo {pages} {1}
  (\bibinfo {year} {1994})}\BibitemShut {NoStop}%
\bibitem [{\citenamefont {Jeffreys}(2000)}]{Jeffreys2000}%
  \BibitemOpen
  \bibfield  {author} {\bibinfo {author} {\bibfnamefont {Harold}\ \bibnamefont
  {Jeffreys}},\ }\href@noop {} {\emph {\bibinfo {title} {Methods of
  mathematical physics}}}\ (\bibinfo  {publisher} {Cambridge university press;
  3rd edition},\ \bibinfo {year} {2000})\BibitemShut {NoStop}%
\bibitem [{\citenamefont {Ballentine}(2014)}]{Ballentine2014}%
  \BibitemOpen
  \bibfield  {author} {\bibinfo {author} {\bibfnamefont {L.}~\bibnamefont
  {Ballentine}},\ }\href@noop {} {\emph {\bibinfo {title} {Quantum mechanics: a
  modern development}}}\ (\bibinfo  {publisher} {World Scientific},\ \bibinfo
  {year} {2014})\BibitemShut {NoStop}%
\bibitem [{\citenamefont {Arfken}\ \emph {et~al.}(2013)\citenamefont {Arfken},
  \citenamefont {Weber},\ and\ \citenamefont {Harris}}]{Arfken2013}%
  \BibitemOpen
  \bibfield  {author} {\bibinfo {author} {\bibfnamefont {George~B.}\
  \bibnamefont {Arfken}}, \bibinfo {author} {\bibfnamefont {Hans~J.}\
  \bibnamefont {Weber}}, \ and\ \bibinfo {author} {\bibfnamefont {Frank~E.}\
  \bibnamefont {Harris}},\ }\href@noop {} {\emph {\bibinfo {title}
  {Mathematical methods for physicists: A comprehensive guide; 7th edition}}}\
  (\bibinfo  {publisher} {Academic press},\ \bibinfo {year} {2013})\BibitemShut
  {NoStop}%
\bibitem [{\citenamefont {Cahill}(2013)}]{Cahill2013}%
  \BibitemOpen
  \bibfield  {author} {\bibinfo {author} {\bibfnamefont {Kevin}\ \bibnamefont
  {Cahill}},\ }\href@noop {} {\emph {\bibinfo {title} {Physical mathematics}}}\
  (\bibinfo  {publisher} {Cambridge University Press},\ \bibinfo {year}
  {2013})\BibitemShut {NoStop}%
\bibitem [{\citenamefont {Wilde}(2017)}]{Wilde2017}%
  \BibitemOpen
  \bibfield  {author} {\bibinfo {author} {\bibfnamefont {Mark}\ \bibnamefont
  {Wilde}},\ }\href@noop {} {\emph {\bibinfo {title} {Quantum information
  theory; 2nd edition}}}\ (\bibinfo  {publisher} {Cambridge University Press},\
  \bibinfo {year} {2017})\BibitemShut {NoStop}%
\bibitem [{\citenamefont {de~Vega}\ and\ \citenamefont
  {Alonso}(2017)}]{Vega2017RMP}%
  \BibitemOpen
  \bibfield  {author} {\bibinfo {author} {\bibfnamefont {Ines}\ \bibnamefont
  {de~Vega}}\ and\ \bibinfo {author} {\bibfnamefont {Daniel}\ \bibnamefont
  {Alonso}},\ }\bibfield  {title} {\enquote {\bibinfo {title} {Dynamics of
  non-markovian open quantum systems},}\ }\href {\doibase
  10.1103/RevModPhys.89.015001} {\bibfield  {journal} {\bibinfo  {journal}
  {Rev. Mod. Phys.}\ }\textbf {\bibinfo {volume} {89}},\ \bibinfo {pages}
  {015001} (\bibinfo {year} {2017})}\BibitemShut {NoStop}%
\bibitem [{\citenamefont {Breuer}\ and\ \citenamefont
  {Petruccione}(2002)}]{BreuerPetruccione2002}%
  \BibitemOpen
  \bibfield  {author} {\bibinfo {author} {\bibfnamefont {{H.-P.}}\ \bibnamefont
  {Breuer}}\ and\ \bibinfo {author} {\bibfnamefont {F.}~\bibnamefont
  {Petruccione}},\ }\href@noop {} {\emph {\bibinfo {title} {The Theory of Open
  Quantum Systems}}}\ (\bibinfo  {publisher} {Oxford University Press},\
  \bibinfo {year} {2002})\BibitemShut {NoStop}%
\bibitem [{\citenamefont {Carmichael}(2003)}]{Carmichael2003}%
  \BibitemOpen
  \bibfield  {author} {\bibinfo {author} {\bibfnamefont {H.~J.}\ \bibnamefont
  {Carmichael}},\ }\href@noop {} {\emph {\bibinfo {title} {Statistical Methods
  in Quantum Optics 1: Master Equations and Fokker-Planck Equations}}}\
  (\bibinfo  {publisher} {Springer},\ \bibinfo {year} {2003})\BibitemShut
  {NoStop}%
\bibitem [{\citenamefont {Weiss}(2012)}]{Weiss2012}%
  \BibitemOpen
  \bibfield  {author} {\bibinfo {author} {\bibfnamefont {U.}~\bibnamefont
  {Weiss}},\ }\href@noop {} {\emph {\bibinfo {title} {Quantum Dissipative
  Systems}}}\ (\bibinfo  {publisher} {World Scientific},\ \bibinfo {year}
  {2012})\BibitemShut {NoStop}%
\bibitem [{\citenamefont {Tong}\ and\ \citenamefont
  {Vojta}(2006)}]{PhysRevLett.97.016802}%
  \BibitemOpen
  \bibfield  {author} {\bibinfo {author} {\bibfnamefont {Ning-Hua}\
  \bibnamefont {Tong}}\ and\ \bibinfo {author} {\bibfnamefont {Matthias}\
  \bibnamefont {Vojta}},\ }\bibfield  {title} {\enquote {\bibinfo {title}
  {Signatures of a noise-induced quantum phase transition in a mesoscopic metal
  ring},}\ }\href {\doibase 10.1103/PhysRevLett.97.016802} {\bibfield
  {journal} {\bibinfo  {journal} {Phys. Rev. Lett.}\ }\textbf {\bibinfo
  {volume} {97}},\ \bibinfo {pages} {016802} (\bibinfo {year}
  {2006})}\BibitemShut {NoStop}%
\bibitem [{\citenamefont {{Strathearn}}\ \emph {et~al.}(2018)\citenamefont
  {{Strathearn}}, \citenamefont {{Kirton}}, \citenamefont {{Kilda}},
  \citenamefont {{Keeling}},\ and\ \citenamefont
  {{Lovett}}}]{Strathearn2018NC}%
  \BibitemOpen
  \bibfield  {author} {\bibinfo {author} {\bibfnamefont {A.}~\bibnamefont
  {{Strathearn}}}, \bibinfo {author} {\bibfnamefont {P.}~\bibnamefont
  {{Kirton}}}, \bibinfo {author} {\bibfnamefont {D.}~\bibnamefont {{Kilda}}},
  \bibinfo {author} {\bibfnamefont {J.}~\bibnamefont {{Keeling}}}, \ and\
  \bibinfo {author} {\bibfnamefont {B.~W.}\ \bibnamefont {{Lovett}}},\
  }\bibfield  {title} {\enquote {\bibinfo {title} {Efficient non-{Markovian}
  quantum dynamics using time-evolving matrix product operators},}\ }\href
  {\doibase 10.1038/s41467-018-05617-3} {\bibfield  {journal} {\bibinfo
  {journal} {Nat. Commun.}\ }\textbf {\bibinfo {volume} {9}},\ \bibinfo {pages}
  {3322} (\bibinfo {year} {2018})}\BibitemShut {NoStop}%
\bibitem [{\citenamefont {J\o{}rgensen}\ and\ \citenamefont
  {Pollock}(2019)}]{Jorgensen2019arxiv}%
  \BibitemOpen
  \bibfield  {author} {\bibinfo {author} {\bibfnamefont {Mathias~R.}\
  \bibnamefont {J\o{}rgensen}}\ and\ \bibinfo {author} {\bibfnamefont
  {Felix~A.}\ \bibnamefont {Pollock}},\ }\bibfield  {title} {\enquote {\bibinfo
  {title} {Exploiting the causal tensor network structure of quantum processes
  to efficiently simulate non-markovian path integrals},}\ }\href {\doibase
  10.1103/PhysRevLett.123.240602} {\bibfield  {journal} {\bibinfo  {journal}
  {Phys. Rev. Lett.}\ }\textbf {\bibinfo {volume} {123}},\ \bibinfo {pages}
  {240602} (\bibinfo {year} {2019})}\BibitemShut {NoStop}%
\bibitem [{\citenamefont {Buser}\ \emph {et~al.}(2017)\citenamefont {Buser},
  \citenamefont {Cerrillo}, \citenamefont {Schaller},\ and\ \citenamefont
  {Cao}}]{Buser_PRA2017_InitialCorrelations}%
  \BibitemOpen
  \bibfield  {author} {\bibinfo {author} {\bibfnamefont {Maximilian}\
  \bibnamefont {Buser}}, \bibinfo {author} {\bibfnamefont {Javier}\
  \bibnamefont {Cerrillo}}, \bibinfo {author} {\bibfnamefont {Gernot}\
  \bibnamefont {Schaller}}, \ and\ \bibinfo {author} {\bibfnamefont {Jianshu}\
  \bibnamefont {Cao}},\ }\bibfield  {title} {\enquote {\bibinfo {title}
  {Initial system-environment correlations via the transfer-tensor method},}\
  }\href {\doibase 10.1103/PhysRevA.96.062122} {\bibfield  {journal} {\bibinfo
  {journal} {Phys. Rev. A}\ }\textbf {\bibinfo {volume} {96}},\ \bibinfo
  {pages} {062122} (\bibinfo {year} {2017})}\BibitemShut {NoStop}%
\end{thebibliography}%

\appendix

\section{Density of states for a bosonic bath}\label{app:bosonDOS}
Consider a collection of non-interacting bosonic modes with Hamiltonian $H = \sum_{k} \omega_{k} b_{k}^{\dagger} b_{k}$.
The partition function of this system takes the form
\begin{equation}
    \ln \mathcal{Z}_{\beta} = - \sum_{k} \ln \left[ 1-e^{-\beta \omega_{k}} \right] .
\end{equation}
In the continuum limit, the sum over modes can be approximated by the integral over a continuous density of modes $g(\omega)$
\begin{equation}
    \ln \mathcal{Z}_{\beta} = - \int_{0}^{\infty} d\omega g(\omega) \ln \left[ 1-e^{-\beta \omega} \right] .
\end{equation}
Expanding the logarithm in powers of $e^{-\beta \omega}$ and re-scaling each term in the resulting series,
we can write the above as
\begin{equation}
    \ln \mathcal{Z}_{\beta} = \int_{0}^{\infty} d\omega \left[ g(\omega) + \frac{g(\omega/2)}{2} + ...  \right] e^{-\beta \omega}
\end{equation}
In the low-temperature limit,
this integral is dominated by the low-energy part of the density of modes.
If we assume that at low-energies the density of modes takes the form $g(\omega) = \alpha \omega^{\gamma}$, where $\alpha$ and $\gamma$ are positive constants,
then the integral takes the form
\begin{equation}
\begin{aligned}
    \ln \mathcal{Z}_{\beta}
    & = \alpha \left( \sum_{n=1}^{\infty}\frac{1}{n^{1+\gamma}} \right) \int_{0}^{\infty} d\omega \omega^{\gamma} e^{-\beta \omega} \\
    & = \alpha \zeta(\gamma+1)\Gamma(\gamma+1) \beta^{-(1+\gamma)} ,
\end{aligned}
\end{equation}
where $\zeta$ denotes the Riemann zeta function and $\Gamma$ the gamma function. Thus
\begin{equation}
    \mathcal{Z}_{\beta} = \exp \left( \alpha \zeta(\gamma+1)\Gamma(\gamma+1) \beta^{-(1+\gamma)} \right) .
\end{equation}
This expression has the general form used in the main text if we make the identification $1+\gamma \rightarrow \gamma$.

\section{Scaling behaviour for the noisy model} \label{app:noise}
In the low-temperature limit, the dominant fine-grained probabilities are those with a vanishing zeroth-order coefficient in the $\small{\text{POVM}}$ energy-gap expansion, and only coarse-grained probabilities containing contributions from such terms are relevant.
For convenience we introduce two sets of fine-grained outcomes: 
First $\Omega_{m} = \{ \mu \, | \, \Delta_{m\mu,0} = \Delta_{m\mu,1} = 0 \}$, which is the set of fine-grained outcomes giving a non-vanishing contribution to the coarse-grained probability of obtaining outcome $m$. Second, $\tilde{\Omega}_{m} = \{ \mu \, | \, \Delta_{m\mu,0} = 0$ and $\Delta_{m\mu,1} \neq 0 \}$, which is the set of fine-grained outcomes for which the contribution to the coarse-grained probability for $m$ vanishes sub-exponentially. Lastly, to simplify the later discussion, we denote the specific outcome within $\tilde{\Omega}_{m}$ which realises the smallest value of the first-order coefficient by $\tilde{\mu}_{m}$.

We now note that if there exists some coarse-grained outcome $m$ such that $\Omega_{m}$ is empty while $\tilde{\Omega}_{m}$ is non-empty, then the arguments presented for the noiseless case also apply to the noisy case, and the same optimal scaling behaviour of the Fisher information can be attained. Thus, in this case, the noise is not detrimental for the scaling. On the other hand, if no such $m$ exists, then we refer to \textit{detrimental noise} (assuming that $\tilde{\Omega}_{m}$ is non-empty for at least one outcome). For detrimental noise we are then left with outcomes for which $\Omega_{m}$ is non-empty,
while $\tilde{\Omega}_{m}$ may or may not be non-empty. We now show that detrimental noise results in a worse scaling compared to the noise-free scenario. This implies that our finite-resolution bound is also applicable to noisy measurements.

Consider the right-hand side of Eq.~\eqref{eq:coarse_fisher_info} for the case of detrimental noise.
For terms where both $\tilde{\Omega}_{m}$ and $\tilde{\Omega}_{n}$ are empty, the scaling behaviour is identical with that of the corresponding noiseless terms (Eq.~\eqref{eq:gs_contribution}), except that the noiseless coefficients of the $\small{\text{POVM}}$ energy gap must be replaced by the coarse-grained version
\begin{equation}
    \Delta_{m,j}^{(c)} \equiv \sum_{\mu\in\Omega_{m}}\frac{p_{m\mu;\beta}}{p_{m;\beta}} \Delta_{m\mu,j} .
\end{equation}
If a coarse-grained second-order $\small{\text{POVM}}$ energy gap exists (that is $\Delta_{m,2}^{(c)}-\Delta_{n,2}^{(c)}\neq 0$ for some $m$ and $n$), then the same scaling behaviour of the Fisher information as given by Eq.~\eqref{eq:gs_contribution} is attainable and this scaling is optimal (note that the probabilities considered here tend to nonzero constants at zero temperature). If a second-order gap does not exist, then the optimal scaling is instead provided by terms for which $\tilde{\Omega}_{m}$ is non-empty for some $m$. A straightforward calculation shows that the contribution from such terms takes the form
\begin{equation} \label{eq:quadratic_noisy_bound}
    \frac{\left[ g_{m\tilde{\mu}_{m}} \Delta_{m\tilde{\mu}_{m},1}\right]^{2}}{\sum_{\mu\in\Omega_{m}}g_{m\mu}} T^{2\Delta_{m\tilde{\mu}_{m},1} - 2} ,
\end{equation}
which should be summed over all outcomes $m$ for which both $\Omega_{m}$ and $\tilde{\Omega}_{m}$ are non-empty.
Assuming that the finite-resolution criterion applies ($\Delta_{m\tilde{\mu}_{m},1}\geq 1$), this contribution is at best constant. Hence under the conditions of finite resolution and detrimental noise, a diverging Fisher information is impossible.

As a second example of a noisy measurement we can consider the coarse-graining of a fine-grained measurement of the form
\begin{equation}
    \begin{aligned}
        \Pi_{00} = \frac{1}{2} e^{-\kappa H} , \ \ \ \ & \Pi_{01} = \frac{1-\eta}{2} e^{-\kappa H}, \\
        \Pi_{10} = \frac{1}{2} \left( \Id - e^{-\kappa H}\right) , \ \ \ \ & \Pi_{11} = \frac{1}{2}\Id - \frac{1-\eta}{2}e^{-\kappa H}.
    \end{aligned}
\end{equation}
This fine-grained model is illustrated in Fig.~\ref{fig:povm_dos}b.
For this measurement we find $\Delta_{00;\beta} = \Delta_{01;\beta} = 0$ and
\begin{equation}
    \begin{aligned}
        & \Delta_{10;\beta} \approx (1+\gamma)T + (1+\gamma) \alpha \kappa T^{2+\gamma} \\
        & \Delta_{11;\beta} \approx (1+\gamma) \frac{\alpha \kappa}{\eta} T^{2+\gamma}.
    \end{aligned}
\end{equation}
Hence, as in the previous example, the fine-grained measurement gives a Fisher information scaling as $T^{\gamma-1}$ to leading order,
and the coarse-grained measurement gives a $T^{2\gamma}$ scaling,
\begin{equation}
    \mathcal{F}_{T} = \frac{(2-\eta) \left( \alpha \kappa \right)^{2}}{\eta}(1+\gamma)^{2} T^{2\gamma} + \mathcal{O}\left(T^{1+3\gamma}\right) .
\end{equation}
Thus the same scaling behaviour of the Fisher information is observed for this alternative example of a noisy model. Note that both models exhibit detrimental noise which results in the different scalings for the fine- and coarse-grained Fisher information.

\section{The non-excitation-preserving interaction as a noisy $\text{\small POVM}$} \label{app:noisy_sbm}
From Eq.~\eqref{eq:prob_approx}, we find that the $\text{\small POVM}$ elements can be written as
\begin{equation}
    \label{eq:apppi1}
    \Pi_1 = t^2\bra{0}H_{\text{int}}\ket{1} \! \bra{1}H_{\text{int}}\ket{0},
\end{equation}
and $\Pi_0=1-\Pi_1$. In the thermal state under consideration, there are no coherences between different bosonic modes and there is no squeezing. Therefore, many terms in Eq.~\eqref{eq:apppi1} do not contribute to the probabilities. Dropping these terms, we can write a slightly simpler $\text{\small POVM}$ that results in the exact same probabilities, capturing the full effect of the measurement
\begin{equation}
    \label{eq:apppi1tilde}
    \tilde{\Pi}_1 = t^{2} \sum_{k} \left( g_{k}^{2} + \lambda_{k}^{2} \right) a^\dagger_k a_k
    + t^{2} \sum_{k} \lambda_{k}^{2},
\end{equation}
and $\tilde{\Pi}_0=1-\tilde{\Pi}_1$. This $\text{\small POVM}$ has an energy gap that has no Taylor expansion, scaling as $T^{2+s}$ in the low temperature limit for $g_k=\lambda_k$ and the spectral density given in Eq.~\eqref{eq:spectraldens}. We can however write the $\text{\small POVM}$ in Eq.~\eqref{eq:apppi1tilde} as a coarse graining over the fine-grained $\text{\small POVM}$ (note the similarity to Eq.~\eqref{eq:exppovmfine})
\begin{equation}
    \label{eq:appfinegr}
    \begin{aligned}
    &\tilde{\Pi}_{11} = \frac{1+\eta}{2}X,\hspace{1.5cm}\tilde{\Pi}_{10} = \frac{1-\eta}{2}(1-X),\\
    &\tilde{\Pi}_{00} = \frac{1+\eta}{2}(1-X),\hspace{0.7cm}\tilde{\Pi}_{10} = \frac{1-\eta}{2}X,
    \end{aligned}
\end{equation}
such that $\tilde{\Pi}_1=\tilde{\Pi}_{11}+\tilde{\Pi}_{10}$ and $\tilde{\Pi}_0=\tilde{\Pi}_{00}+\tilde{\Pi}_{01}$.
Here we introduced
\begin{equation}
    \eta = 1-2 t^2\sum_k\lambda_k^2,
\end{equation}
and
\begin{equation}
    X = \frac{ t^2\sum_k(g_k^2+\lambda_k^2)a_k^\dagger a_k}{1-2 t^2\sum_k\lambda_k^2}.
\end{equation}
The fine-grained $\text{\small POVM}$ elements are of the same form as the $\text{\small POVM}$ elements for the excitation-preserving case. Indeed, setting $\lambda_k=0$, only $\tilde{\Pi}_{00}$ and $\tilde{\Pi}_{11}$ remain finite but do not change their form. We therefore find the same $\text{\small POVM}$ gaps as for the excitation-preserving case
\begin{equation}
    \label{eq:appgapsfinegr}
    \Delta_{00} = \Delta_{10} = 0,\hspace{1cm}\Delta_{11}=\Delta_{01}=(1+s)T.
\end{equation}
The Fisher information for the fine-grained $\text{\small POVM}$ thus scales as $T^{s-1}$. The coarse grained $\text{\small POVM}$ gap is determined by Eq.~\eqref{eq:gap_cg} and reads
\begin{equation}
    \label{eq:appgapcg}
    \Delta_1 = \frac{p_{11}}{p_{11}+p_{10}}(1+s)T,
\end{equation}
which scales as $T^{s+2}$ for the scenario considered in the main text.

\section{Tensor network simulation}\label{app:tensor} \label{app:tempo}
Here we provide details of the numerical methods behind the result shown in \figref{fig:FI_sbm}. We consider the ground state probability
\begin{equation}
    p_{0;\beta}^{(k)} = \tr \left\lbrace \hat{P}_{0} \ \mathcal{U}_{\delta t}^{k}
    \left[ \hat{P}_{0} \otimes \rho_{\beta} \right] \right\rbrace ,
\end{equation}
where $\hat{P}_{0}$ is a projection operator onto the qubit ground state $\ket{0}$,
and we have decomposed the unitary evolution into $k$-steps of duration $\delta t$.
Furthermore we consider the spin-boson model
\begin{equation}
    \hat{H} = \sum_{k}\omega_{k} \hat{a}^{\dagger}_{k} \hat{a}_{k} + \frac{1}{2} \Omega \hat{\sigma}_{z}
    + \frac{1}{2}\hat{\sigma}_{x} \sum_{k} g_{k} \left( a_{k} + a_{k}^{\dagger} \right) .
\end{equation}
The spin-boson model can be numerically simulated using recently developed tensor network methods \cite{Strathearn2018NC,Jorgensen2019arxiv}. Taking each unitary step to be of a short duration we can make the approximation (Trotter-Suzuki decomposition)
\begin{equation}
    U_{\delta t} = W_{\delta t/2} V_{\delta t} W_{\delta t/2} + \mathcal{O}(\delta t^{3}) ,
\end{equation}
where $W_{\delta t} = \exp \left(-i \delta t (\hat{H}-\Omega \hat{\sigma}_{z}/2)\right)$ describes the influence of the sample on the probe qubit, and $V_{\delta t} = \exp \left(-i \delta t \Omega \hat{\sigma}_{z}/2 \right)$ describes the free evolution of the probe qubit. As the interaction term is diagonal in the eigenstates of the operator $\hat{\sigma}_{x}$, we can expand the ground state probability in terms of these eigenstates. This gives rise to a discrete Feynman-Vernon Influence functional, which can be summed analytically. The ground state probability then takes the form
\begin{equation}
\begin{aligned}
    p_{0;\beta}^{(k)} = \sum_{\left\lbrace \alpha \right\rbrace} \ 
    & \hat{P}_{0}^{\alpha_{2k+1}} \mathcal{V}_{\delta t}^{\alpha_{2k}\alpha_{2k+1}} ... \mathcal{V}_{\delta t}^{\alpha_{2}\alpha_{1}} \hat{P}_{0}^{\alpha_{0}} \\
    & \times \left[ \Pi_{i=1}^{2k} \Pi_{j=1}^{i} \mathcal{A}_{\beta}^{\alpha_{i}\alpha_{j}} \right]
    \left[ \Pi_{l=0}^{k} \delta_{\alpha_{2l+1},\alpha_{2l}} \right] \ .
\end{aligned}
\end{equation}
where we have introduced a compound index $\alpha = (s,r)$ of spin-x eigenvalues, $\delta_{\alpha_{i},\alpha_{j}}$ denotes the Kronecker delta function,
$\hat{P}_{0}^{\alpha} = \bra{s} \! \hat{P}_{0} \! \ket{r}$,
and $\mathcal{V}$ are the Liouville operators representing the free dynamics of the ancilla qubit
\begin{equation}
    \mathcal{V}_{\delta t}^{\alpha \alpha'} = \bra{s} V_{\delta t} \ket{s'}\! \bra{r'} V_{\delta t}^{\dagger} \ket{r}.
\end{equation}
The influence tensors, $\mathcal{A}_{\beta}^{\alpha_{i}\alpha_{j}}$, describe the influence of the sample on the state of the qubit and contain all the temperature dependence of the probability. For linearly coupled models, the individual tensors depend only on the time separation $(i-j) \delta t/2$. The influence tensors are given by
\begin{equation}
	\mathcal{A}_{\beta}^{\alpha_{i} \alpha_{j}} = e^{ -(s_{i}-r_{i})( \eta_{i-j}s_{j}-\eta_{i-j}^{*}r_{j} )},
\end{equation}
expressed in terms of the memory kernel elements
\begin{equation}
	\eta_{i-j} = 
	\begin{cases}
	\int_{t_{i-1}}^{t_{i}} \int_{t_{j-1}}^{t_{j}} dt' dt'' C(t'-t'') \ ,& \ \ i \neq j \\
	\int_{t_{i-1}}^{t_{i}} \int_{t_{i-1}}^{t'} dt' dt'' C(t'-t'') \ ,& \ \ i = j
	\end{cases},
\end{equation}
which are themselves defined in terms of the bath auto-correlation function
\begin{align}
	C(t) = \frac{1}{\pi}\int_{0}^{\infty} d\omega \rho(\omega) \frac{\cosh \left[ \omega(\beta-it) \right]}{\sinh \left( \beta \omega/2 \right)}.
\end{align}
The bath auto-correlation function is given in terms of the spectral density $\rho(\omega)$ introduced in the main text.

The attainable temperature estimation precision depends not only on the ground state probability, but also on the derivative of this probability.
Computing the derivative of the distribution with respect to the inverse temperature gives
\begin{equation}
\begin{aligned}
    \partial_{\beta} p_{0\beta}^{(k)} 
    = & \sum_{i=1}^{2k}\sum_{j=1}^{i} \mu_{ij} \sum_{\left\lbrace \alpha \right\rbrace} \left[ \Pi_{l=0}^{k} \delta_{\alpha_{2l+1},\alpha_{2l}} \right] \\
    & \times \hat{P}_{0}^{\alpha_{2k+1}} \mathcal{V}_{\delta t}^{\alpha_{2k}\alpha_{2k+1}} ... \mathcal{V}_{\delta t}^{\alpha_{2}\alpha_{1}}
    \hat{P}_{0}^{\alpha_{0}} \\
    & \ \ \times \left[ \Pi_{i=1}^{2k} \Pi_{j=1}^{i} \mathcal{A}_{\beta}^{\alpha_{i}\alpha_{j}} \right]
    \alpha^{-}_{i}\alpha^{-}_{j}\\
\end{aligned}
\end{equation}
where we have defined $\alpha^{-} = s-r$.
It turns out that the same tensor network methods used to compute the probability can be used to compute the derivative of the probability. Furthermore we have defined $\mu_{ij} = - \partial_{\beta} \eta_{i-j}$, the square of which gives the Fisher information scaling at low-temperatures.
At low temperatures, all the temperature dependence of the ground-state probability comes from these coefficients. We can approximate them by the series
\begin{equation}
\begin{aligned}
    \mu_{ij} = 
    & \frac{\alpha \delta t^{2}}{4 \beta^{\gamma+2}} \times \left[\Gamma(\gamma+2) 
    - \frac{\delta t^{2}}{8\beta^{2}}(i-j)^{2} \Gamma(\gamma+4) \right. \\
    & \left. + \frac{\delta t^{4}}{376\beta^{4}}(i-j)^{4} \Gamma(\gamma+6) - ... \right]
\end{aligned}
\end{equation}
This shows that, to leading order, the exact expressions reproduce the low-temperature Fisher-information scaling obtained within the short-time approximation.

\end{document}